\documentclass[11pt,a4paper]{article}
\usepackage{jheppub_kim}
\usepackage{rotating} 
\usepackage{graphicx,epsfig}
\usepackage{amsmath}
\usepackage {amssymb}
\usepackage{subfigure}

\usepackage{relsize}

\newcommand{\be}{\begin{equation}}
\newcommand{\ee}{\end{equation}}
\newcommand{\bea}{\begin{eqnarray}}
\newcommand{\eea}{\end{eqnarray}}
\newcommand{\beaa}{\begin{eqnarray*}}
\newcommand{\eeaa}{\end{eqnarray*}}

\newcommand{\nn}{\nonumber \\}
\newcommand{\e}{\mathrm{e}}

\begin{document}

\title{Singular cosmological evolution using canonical and ghost scalar fields}

\author[a,b]{Shin'ichi Nojiri}

\author[c,d]{S.D. Odintsov}

\author[e,f,g]{V.K. Oikonomou}

\author[h,i]{Emmanuel N. Saridakis}

\affiliation[a]{Department of Physics, Nagoya University, Nagoya 464-8602, Japan}

\affiliation[b]{Kobayashi-Maskawa Institute for the Origin of Particles and the
Universe, Nagoya University, Nagoya 464-8602, Japan}

\affiliation[c]{Institut de Ciencies de l'Espai (IEEC-CSIC), Campus UAB,
Torre C5-Par-2a pl, E-08193 Bellaterra, Barcelona, Spain}

\affiliation[d]{InstituciŽo Catalana de Recerca i Estudis Avanžcats (ICREA), Barcelona, 
Spain}

\affiliation[e]{Department of Theoretical Physics, Aristotle University of Thessaloniki,
54124 Thessaloniki, Greece}

\affiliation[f]{National Research Tomsk State University, Tomsk, 634061 Russia}

\affiliation[g]{Tomsk State Pedagogical University, Tomsk, 634061 Russia}

\affiliation[h]{Physics Division, National Technical University of Athens,
15780 Zografou Campus,  Athens, Greece}

\affiliation[i]{Instituto de F\'{\i}sica, Pontificia
Universidad de Cat\'olica de Valpara\'{\i}so, Casilla 4950,
Valpara\'{\i}so, Chile}

\emailAdd{nojiri@gravity.phys.nagoya-u.ac.jp}
\emailAdd{odintsov@ieec.uab.es}
\emailAdd{v.k.oikonomou1979@gmail.com}
\emailAdd{Emmanuel$_-$Saridakis@baylor.edu}

\abstract{We demonstrate that finite time singularities of Type IV can be consistently 
incorporated in the Universe's cosmological evolution, either appearing in the 
inflationary era, or in the late-time regime. While using only one scalar 
field instabilities can in principle occur at the time of the phantom-divide 
crossing, when two fields are involved we are able to avoid such instabilities. 
Additionally, the two-field scalar-tensor theories prove to be able to offer a plethora 
of possible viable cosmological scenarios, at which various types of cosmological 
singularities can be realized. Amongst others, it is possible to describe inflation with 
the appearance of a Type IV singularity, and phantom late-time acceleration which ends in 
a Big Rip. Finally, for completeness, we also present the Type IV realization 
in the context of suitably reconstructed $F(R)$ gravity.}


\maketitle

\section{Introduction}

The observational data concerning the effective equation-of-state (EoS) parameter 
of dark energy $w_{\mathrm{eff}}$, slightly indicate that in the recent cosmological 
past the EoS parameter might have crossed the phantom divide and lies between 
quintessence 
and phantom regimes \cite{Nojiri:2005pu,ref2,Jassal:2005qc,Cai:2009zp}, and therefore the 
cosmological evolution of the present Universe might be described by an effective phantom 
phase. There might also be the possibility that although the current Universe may be of 
quintessential type, the phantom era may occur in the near future. For an incomplete list 
of articles discussing the possibility that our Universe is described by an effective 
late-time phantom era see 
\cite{Nojiri:2005pu,ref2,Jassal:2005qc,Cai:2009zp,Starobinsky:1999yw,
Elizalde:2004mq,Onemli:2004mb,
Caldwell:1999ew,
Faraoni:2005gg,Nojiri:2003jn,ref3,Singh:2003vx,Sami:2003xv, 
Guo:2004fq,Saridakis:2009uu,GonzalezDiaz:2003rf} 
and references therein. On the other hand, the inflationary era is also an accelerating 
phase of the Universe, which is not up to date excluded to be of quintessence or phantom 
type. One of the biggest problems in this type of phenomenology, is to fully understand 
the relation and the connection of early-time inflation with late-time acceleration, and 
in particular how to consistently describe these accelerating eras using the same 
theoretical framework. The origin of the problem itself might be the incomplete 
understanding we have, up to date, of dark energy and how the transition from 
deceleration to the dark energy acceleration epoch is eventually achieved.

As was demonstrated in \cite{Nojiri:2005pu}, a promising theoretical unified description 
of both early-time phantom inflation and late-time phantom acceleration is achieved by 
using canonical and non-canonical scalar fields. Using this generalized 
multiple-scalar-field scalar-tensor theory, with generalized scalar-field-dependent 
kinetic functions in front of the kinetic terms and with a generalized scalar potential, 
a consistent theoretical framework can be provided for these theories 
\cite{Boisseau:2000pr}. This can be done 
either by using two scalar fields, or by using even a single scalar field. However, in 
the latter case there is a theoretical obstacle connected with the transition time at 
which the EoS evolves from non-phantom to phantom and vice-versa. In particular, the 
single scalar field theoretical description of cosmological dynamics always comes with 
the cost of possible instabilities at the cosmic time that the phantom divide is crossed, 
since at exactly this point the dynamical system that describes the evolution of the 
Universe can be unstable towards linear perturbations \cite{vikman}, rendering the 
resulting scalar-tensor theory unstable. Hence, the use of two scalar fields is necessary 
in order to resolve these issues in a concrete way \cite{Nojiri:2005pu,Ito:2011ae}. 

Definitely, the single-field scalar-tensor theory can in some cases prove valuable, 
since it offers the opportunity to describe exotic cosmological scenarios, which were 
not possible to be realized in the context of standard Einstein-Hilbert general 
relativity. In a recent study \cite{Nojiri:2015fra} we addressed the issue of having an 
inflationary era in the presence of finite-time singularities, with special emphasis in 
the Type IV singularity. Particularly, as we explicitly have shown, it is possible to 
have viable phenomenology even in the presence of Type IV singularities, and in some 
cases compatible with observational data coming from Planck \cite{planck} and BICEP2 
\cite{BICEP2} collaborations. These finite-time singularities can be of various types and 
were classified in \cite{Nojiri:2005sx}, with regards to their relevance in the 
cosmological evolution. These types vary, with the main criterion that classifies them 
being the fact that certain observable quantities may become infinite at the time these 
occur. 

In principle, a singularity in a cosmological context is an unwanted feature of 
cosmological evolution. Space-time singularities were studied long ago by Hawking and 
Penrose \cite{hawkingpenrose}, and they are defined as isolated points in space-time at 
which the space-time is geodesically incomplete. This feature practically means that 
there exist null and time-like geodesics, which cannot be continuously extended to 
arbitrary values of their parameters. The most characteristic example of this type of 
singularity is the initial singularity of our Universe, which naturally appears in 
inflationary theories \cite{inflation,Gorbunov:2011zzc,Linde2014,Lyth:1998xn}. It is 
worthy to mention however that there exist theories which are free of this initial 
singularity, such as the bouncing Universe scenarios 
\cite{Starobinskii1978,Novello:2008ra,Brandenberger:1993ef,bounce3,bounce5,
Qiu:2013eoa,Cai:2014xxa,Khoury:2001zk,Erickson:2003zm, 
Lehners:2013cka,Saridakis:2007cf,Creminelli:2007aq,Cai:2010zma, HLbounce,
Bojowald:2001xe,Martin:2003sf,Cai:2012ag,Ashtekar:2007tv,
Cailleteau:2012fy,Haro:2014wha,Amoros:2014tha,WilsonEwing:2012pu}.
Apart from these severe singularities however, there exist singularities at which some 
observable quantities may become infinite, without this implying geodesic incompleteness. 
These are called sudden singularities and were 
extensively studied by Barrow \cite{barrowsing2} (see also 
\cite{Shtanov:2002ek,Gorini:2003wa,barrowsing3,Lake:2004fu,Cotsakis:2004ih,
Dabrowski:2004bz,
FernandezJambrina:2004yy,Barrow:2004he,Nojiri:2004ip,
Stefancic:2004kb,sergnojnew1,Nojiri:2005sr,Cattoen:2005dx,Sami:2006wj,
BouhmadiLopez:2006fu,
Yurov:2007tw,FernandezJambrina:2008dt,sergbam08,Barrow:2009df,
BouhmadiLopez:2009pu, Bamba:2009uf, Barrow:2010wh,Barrow:2015ora} for some later 
studies). 
In these singularities, the scale factor remains finite at the singular point, and 
therefore these are not singularities of crushing type, such as the Big Rip 
\cite{Nojiri:2005sx,Caldwell:2003vq,ref5,Faraoni:2001tq,Gorini:2002kf,
Singh:2003vx,Chimento:2003qy,Stefancic:2003rc,Sami:2003xv,GonzalezDiaz:2003rf,
Chimento:2004ps,Csaki:2004ha,Hao:2004ky,Wu:2004ex,Dabrowski:2004hx,Aref'eva:2004vw,
Zhang:2005eg,Elizalde:2005ju,Cai:2005ie,Lobo:2005us,Sola:2005et,
Barbaoza:2006hf}. 

In our recent study \cite{Nojiri:2015fra}, the focus was on the existence of a 
non-crushing type of singularity, namely the Type IV. Our theoretical description of the 
cosmological evolution involved a scalar-tensor theory with a single scalar field. 
However, when someone uses a single scalar field, the scalar-tensor description might be 
unstable at the cosmic time at which the phantom divide is crossed. This was not the case 
in our previous article, but this is a clear possibility when someone deals with 
single-field scalar-tensor theories. Therefore, the purpose of the present article is 
two-fold. Firstly, we aim to highlight the problem of instability in the single-field 
scalar-tensor cosmologies, in the presence of finite time singularities. Secondly, 
we are interested to formally address the crossing from the non-phantom to the 
phantom era consistently. The latter issue is definitely connected to the first one. We 
shall exemplify the problem by using a single-field scalar-tensor theory to produce a 
variant of a well known potential from inflationary cosmology, namely the hilltop 
potential 
\cite{hilltop,Linde:1981mu,Kohri:2007gq,Dutta:2008qn,Dutta:2009yb,Ijjas:2013vea,
Antusch:2014qqa,Boubekeur:2005zm}, in which we consistently incorporate a Type IV 
singularity. As we show, the solution is unstable at the time the EoS crosses the phantom 
divide, and consequently the two-scalar-field cosmological description is rendered 
compelling. Having done this, we shall study illustrative examples of two-field 
scalar-tensor cosmologies, in which finite-time singularities exist. More importantly, we 
shall explicitly demonstrate that it is possible to unify phantom (or non-phantom) 
inflation, with phantom (or non-phantom) late-time acceleration. Finally, for 
completeness, we shall investigate which $F(R)$ gravity can generate a cosmological 
evolution corresponding to the hilltop scalar-tensor model, near the Type IV singularity. 
Interestingly enough, if the Type IV singularity is assumed to occur 
at late times, the resulting $F(R)$ gravity is of the form $R+\frac{A}{R}$, a well known 
model that can consistently describe late-time acceleration \cite{Carroll:2003wy}.

This paper is organized as follows: In section \ref{genanalysis} we show that the 
single-field scalar-tensor theory may lead to instabilities at the phantom-divide 
crossing, which are removed with the use of a second field. In section 
\ref{variousexamples} we proceed to the investigation of various specific cosmological 
evolutions in the framework of two-field scalar-tensor theory, which prove the 
capabilities of the construction. In particular, in subsection \ref{subgenexamples} we 
reconstruct scenarios which exhibit Type II or Type IV singularities, in subsection  
\ref{subslowroll} we analyze the slow-roll inflationary realization in such theories, in 
\ref{subsingDE} we reconstruct a universe with a singular inflationary phase that 
results in a dark energy epoch which ends in a Big Rip, in \ref{subBarrow} we present the 
two-field Barrow's model, while in subsection \ref{subsmultiple} we extend our 
investigation in the case of multiple scalar fields. In section \ref{FRsection}, for 
completeness, we also present the Type IV realization in the context of suitably 
reconstructed $F(R)$ gravity. Finally, in section \ref{conclus} we summarize our results.

\section{Scalar-tensor gravity: general analysis}
\label{genanalysis}

In this section we shall examine in detail how the finite-time singularities can be 
consistently incorporated in generic scalar-tensor theories containing two scalar fields. 
In addition, we will demonstrate why the need for two scalars is in some cases 
compelling. Using well known reconstruction methods 
\cite{Nojiri:2005pu,Capozziello:2005tf}, for a given cosmological evolution we shall 
provide the details of the generic scalar-tensor models that can consistently incorporate 
finite-time singularities, and then in the next section we will explicitly demonstrate 
our results using some illustrative examples. We mention to the reader that apart from 
Refs. 
\cite{Nojiri:2005pu,Capozziello:2005tf}, many reconstruction techniques had been 
developed earlier, and 
most of 
these studies are very well reviewed in Ref. \cite{sahnistaro}.

In this work we consider a spatially flat Friedmann-Robertson-Walker (FRW) metric 
of the form 
\be
\label{JGRG14}
ds^2 = - dt^2 + a(t)^2 \sum_{i=1,2,3} \left(dx^i\right)^2\, ,
\ee
with $a(t)$ the scale factor. In this case, the effective energy density 
$\rho_\mathrm{eff} $ and the effective pressure
$p_\mathrm{eff}$ are defined as 
\be
\label{IV}
\rho_\mathrm{eff} \equiv \frac{3}{\kappa^2} H^2 \, , \quad
p_\mathrm{eff} \equiv - \frac{1}{\kappa^2} \left( 2\dot H + 3 H^2
\right)\, ,
\ee
where $H=\dot{a}/a$ denotes the Hubble parameter and  dots indicate differentiation 
with respect to the cosmic time $t$.

Let us first recall the finite-time cosmological singularities classification. For 
details 
on this subject the reader is referred to 
\cite{Nojiri:2005sx,Nojiri:2006ri,
delaCruzDombriz:2012xy,sergnoj}. According 
to \cite{Nojiri:2005sx,sergnoj}, the finite-time future singularities are classified in 
the following way: 
\begin{itemize}

\item Type I (``Big Rip'') : As $t \to t_s$, the scale factor $a$, the effective energy 
density $\rho_\mathrm{eff}$, and the effective pressure $p_\mathrm{eff}$ diverge, namely 
$a \to \infty$, $\rho_\mathrm{eff} \to \infty$, and
$\left|p_\mathrm{eff}\right| \to \infty$.
For a presentation of this type of singularity, see Ref.~\cite{Caldwell:2003vq} and also 
Ref.~\cite{Nojiri:2005sx}.

\item Type II (``sudden''): As $t \to t_s$, both the scale factor and the effective 
energy 
density are finite, that is $a \to a_s$, $\rho_\mathrm{eff} \to 
\rho_s$, but the effective pressure diverges, namely $\left|p_\mathrm{eff}\right| \to 
\infty$ 
\cite{barrowsing2,Shtanov:2002ek,Gorini:2003wa,barrowsing3,
Lake:2004fu,Cotsakis:2004ih,
Dabrowski:2004bz,
FernandezJambrina:2004yy,Barrow:2004he,Nojiri:2004ip,
Stefancic:2004kb,sergnojnew1,Nojiri:2005sr,Cattoen:2005dx,Sami:2006wj,
BouhmadiLopez:2006fu, Yurov:2007tw,
FernandezJambrina:2008dt,sergbam08,Barrow:2009df,BouhmadiLopez:2009pu,Bamba:2009uf,
Barrow:2010wh}.

\item Type III: As $t \to t_s$, the scale factor is finite, $a \to a_s$ but the  
effective energy density and the effective pressure diverge, $\rho_\mathrm{eff} \to 
\infty$, $\left|p_\mathrm{eff}\right| \to \infty$ \cite{Nojiri:2005sx}.

\item Type IV: As $t \to t_s$, the scale factor, the effective energy density, and
the effective pressure are finite, namely $a \to a_s$, $\rho_\mathrm{eff} \to \rho_s$,
$\left|p_\mathrm{eff}\right| \to p_s$, but the Hubble's rate higher derivatives diverge, 
that is $H\equiv \dot a/a$ \cite{Nojiri:2005sx}.
\end{itemize}

Finally, note that in the aforementioned list of singularities, once should add the 
standard case of recollapse due to a positive spatial curvature ending in the ``Big
Crunch'' \cite{BarrowGalloway}.

In a previous work \cite{Nojiri:2015fra} we studied how singular inflation can be 
properly accommodated to a power-law model of inflation, studied by Barrow in 
\cite{Barrow:2015ora}. Using the scalar field reconstruction method 
\cite{Nojiri:2005pu,Capozziello:2005tf}, we found which scalar-tensor gravities can lead 
to a successful description of such inflationary cosmology. However, the conventions we 
used in the previous article lead to a stable scalar-tensor solution, which had an 
effective equation of state characteristic of quintessential acceleration, as long as 
the power-law inflation model is considered. Nevertheless, the stability of the 
scalar-tensor solutions is not ensured in general, and instabilities might occur during 
the reconstruction process. Therefore, in this section we explicitly demonstrate how 
instabilities can indeed appear, by using two phenomenologically appealing inflationary 
models. As we evince, the simplest way to avoid the instabilities is to use two (or more) 
scalar fields, with one of them being ghost and the other scalar field being 
canonical.

In a first subsection we aim to show how the Type IV singularities can occur in a 
slightly deformed version of a very well known viable inflationary model, namely the 
hilltop inflationary potential 
\cite{hilltop,Linde:1981mu,Kohri:2007gq,Dutta:2008qn,Dutta:2009yb,Ijjas:2013vea,
Antusch:2014qqa,Boubekeur:2005zm}. Thus, in a next subsection we will be able to provide 
a consistent framework consisting of two scalar fields for the aforementioned 
inflationary 
cosmologies.

\subsection{Singular evolution with hilltop-like potentials: Description with a single 
scalar and the singularity appearance}

We start our analysis with the hilltop potentials, and we demonstrate how a Type IV 
singularity can be consistently incorporated in the theoretical framework of these 
models by using a very general reconstruction scheme for scalar-tensor theories. Before 
getting into the main analysis, it is worthy to give a brief description of the scalar 
reconstruction method we shall use. For a detail presentation on this issue the reader is 
referred to \cite{Nojiri:2005pu,Capozziello:2005tf}. 

We consider the following non-canonical scalar-tensor action, which describes a single 
scalar field:
\be
\label{ma7}
S=\int d^4 x \sqrt{-g}\left\{
\frac{1}{2\kappa^2}R - \frac{1}{2}\omega(\phi)\partial_\mu \phi
\partial^\mu\phi - V(\phi) + L_\mathrm{matter} \right\}\, ,
\ee
with the function $\omega(\phi)$ being the kinetic function and $V(\phi)$  the scalar 
potential. The kinetic-term function $\omega(\phi)$ appearing in  (\ref{ma7}) is 
irrelevant, since it can be consistently absorbed by appropriately redefining the 
scalar field $\phi$ as 
\be
\label{ma13}
\varphi \equiv \int^\phi d\phi \sqrt{\omega(\phi)} \,,
\ee
where it is assumed that $\omega(\phi)>0$. Hence, the kinetic term of the scalar field 
appearing in action (\ref{ma7}) can be written as
\be
\label{ma13b}
 - \omega(\phi) \partial_\mu \phi \partial^\mu\phi
= - \partial_\mu \varphi \partial^\mu\varphi\, .
\ee
On the other hand, if  $\omega(\phi)<0$ then the scalar field becomes ghost, and it can 
describe the phantom dark energy. In this case, instead of (\ref{ma13}) one makes the 
redefinition 
\be
\label{ma13p}
\varphi \equiv \int^\phi d\phi \sqrt{-\omega(\phi)} \, ,
\ee
and thus, instead of (\ref{ma13b}), one acquires the expression
\be
\label{ma13bp}
 - \omega(\phi) \partial_\mu \phi \partial^\mu\phi
= \partial_\mu \varphi \partial^\mu\varphi\, .
\ee
When the scalar field is a ghost field the energy density corresponding to the 
classical theory becomes unbounded from below. At the quantum level the energy 
could be rendered bounded from below, however at the significant cost of the 
appearance of a negative norms \cite{Cline:2003gs,Nojiri:2013ru}.

In both canonical and phantom cases, in the case of FRW geometry the energy density and 
pressure of the scalar field write as 
\be
\label{ma8}
\rho = \frac{1}{2}\omega(\phi){\dot \phi}^2 + V(\phi)\, ,\quad
p = \frac{1}{2}\omega(\phi){\dot \phi}^2 - V(\phi)\, .
\ee
Having these at hand, we can use the usual Friedmann equations in the absence of the 
matter sector, in order to promptly express the scalar potential $V(\phi)$ and the 
kinetic term $\omega(\phi)$ in terms of the Hubble rate and it's first derivative as
\be
\label{ma9}
\omega(\phi) {\dot \phi}^2 = - \frac{2}{\kappa^2}\dot H\, ,\quad
V(\phi)=\frac{1}{\kappa^2}\left(3H^2 + \dot H\right)\, .
\ee
The scalar-reconstruction method is based on the assumption that the kinetic term 
$\omega(\phi)$ and the scalar potential $V(\phi)$, can be written in terms of a single 
function $f(\phi)$ in the following way:
\be
\label{ma10}
\omega(\phi)=- \frac{2}{\kappa^2}f'(\phi)\, ,\quad
V(\phi)=\frac{1}{\kappa^2}\left[3f(\phi)^2 + f'(\phi)\right]\, .
\ee
Therefore, the equations (\ref{ma9}) become \be
\label{ma11}
\phi=t\, ,\quad H=f(t)\, .
\ee
Finally, by varying the action in terms of the scalar field we obtain its evolution 
equation as
\be
\label{ma12}
0=\omega(\phi)\ddot \phi + \frac{1}{2}\omega'(\phi){\dot\phi}^2
+ 3H\omega(\phi)\dot\phi + V'(\phi)\, ,
\ee
which is also satisfied by the solution (\ref{ma11}).  

In order to proceed, we need to make a choice for the scalar potential. In the case of 
a canonical scalar field, the hilltop inflation potential 
\cite{hilltop,Linde:1981mu,Kohri:2007gq,Dutta:2008qn,Dutta:2009yb,Ijjas:2013vea,
Antusch:2014qqa,Boubekeur:2005zm} has been proved 
to be consistent with the Planck data \cite{planck}, and it is approximately given by
\begin{equation}
\label{hilltop}
V(\varphi )\simeq \mathcal{A} -\mathcal{B}\varphi^p+\ldots \, ,
\end{equation}
with the ellipsis denoting higher-order terms which are negligible during inflation. When 
the parameters are chosen appropriately, and amongst others $p=2$, the implications of 
this inflationary mode are in $95 \%$ agreement with the Planck data \cite{planck}. 
Hence, we prefer to consider a case close to this phenomenologically well-behaved case, 
and we assume $p\simeq 2-\epsilon$, with $\epsilon$ an infinitesimally small number 
$|\epsilon| \ll 1$. 

Let us now start the main part of the analysis of this subsection, which is to examine 
whether there exist finite-time singularities in the hilltop potential scenario 
(\ref{hilltop}). As we shall demonstrate, it is possible to connect a Type IV and a Type 
I singularity with potentials having a functional resemblance to the hilltop potential. 
However, it is impossible to fully satisfy the constraints posed by Planck in these 
hilltop-like potentials, and therefore we should stress that the following considerations 
are excluded by observations. Nevertheless, for the moment we are interested in 
demonstrating that instabilities will appear, and we do not consider consistent 
confrontation with observations. One could construct more complicated single-field 
scenarios, in agreement with observations, and with the appearance of singularities, 
however this would lie beyond our scope, which is to show in a simple single-field model 
that singularities appear. 

In order to quantify the appearance of a singularity we parameterize the Hubble rate as
\be
\label{IV1}
H(t) = c_0 + b_0 \left( t_s - t \right)^\alpha\, ,
\ee
where $c_0,b_0$ are arbitrary positive constants, and later on we shall determine which 
values of $\alpha$ are allowed. Clearly, according to the value of $\alpha$, one could 
have the appearance of the following singularities:
\begin{itemize}
\label{lista}
\item $\alpha<-1$ corresponds to Type I singularity.
\item $-1<\alpha<0$ corresponds to Type III singularity.
\item $0<\alpha<1$ corresponds to Type II singularity.
\item $\alpha>1$ corresponds to Type IV singularity.
\end{itemize}
Inserting the above Hubble ansatz into (\ref{ma10}), we may directly find the scalar 
potential and the kinetic term as
\begin{align}
\label{IV3}
\omega(\phi) = & \frac{2 b_0 \alpha  (t_s-\phi )^{-1+\alpha }}{\kappa ^2}\, ,\\
V(\phi) = & \frac{3 c_0^2}{\kappa ^2}-\frac{b_0 \alpha  (t_s-\phi )^{-1+\alpha }}{\kappa 
^2}+\frac{
6 b_0 c_0 (t_s-\phi )^{\alpha }}{\kappa ^2}+\frac{3 b_0^2 (t_s-\phi )^{2 \alpha }}{\kappa 
^2}\, .
\label{IV3b}
\end{align}
Additionally, imposing the transformation (\ref{ma13}) we obtain 
\be
\label{IV5}
\varphi = - \frac{2 \sqrt{2\alpha b_0}}{\kappa \left(\alpha+ 1 \right) }
\left( t_s - \phi \right)^{\frac{\alpha + 1}{2}}\, ,
\ee
and therefore the scalar potential (\ref{IV3b}) becomes
\be
\label{IV6}
V(\varphi) \sim \frac{3 c_0^2}{\kappa ^2}- \frac{b_0 \alpha  
\mathcal{A}_1^{\alpha-1}}{\kappa ^2}\varphi^{\frac{2(-1+\alpha)}{\alpha+1}}+\frac{6 b_0 
c_0 \mathcal{A}_1^{\alpha }}{\kappa ^2}\varphi^{\frac{2\alpha}{\alpha+1}}+\frac{3 b_0^2 
\mathcal{A}_1^{2 \alpha }}{\kappa ^2}\varphi^{\frac{4\alpha}{
\alpha+1} }\, ,
\ee
with  
\begin{equation}\label{a1}
\mathcal{A}_1=\left [- \frac{2 \sqrt{2\alpha b_0}}{\kappa \left(\alpha+ 1 \right) } 
\right]^{\frac{2}{\alpha+1}}\, .
\end{equation}

Since we are interested in small values of $\varphi$, which according to (\ref{ma11}) 
corresponds to the case where $t_s-\phi\rightarrow 0$, we may neglect the higher-order 
terms in the potential (\ref{IV6}), and approximate it as
\be
\label{IV6a}
V(\varphi) \approx \frac{3 c_0^2}{\kappa ^2}- \frac{b_0 \alpha  \mathcal{A}_1^{-1+\alpha 
}}{\kappa ^2}\varphi^{\frac{2(\alpha-1)}{\alpha+1} }\, .
\ee
In the case $p=2-\epsilon$ this potential coincides with (\ref{hilltop}) under the 
identification
\begin{equation}\label{eqnconstr1}
\alpha=\frac{2+\epsilon}{\epsilon}\, ,
\end{equation}
which is a very large positive number as $\epsilon\rightarrow 0$, if $\epsilon>0$. Hence, 
in this case the system develops a Type IV singularity. On the other hand, when
$p>2$ the system develops a Type I singularity, since in this case comparing  
(\ref{hilltop}) with (\ref{IV6a}) we obtain 
\begin{equation}
\label{eqnconstr2B}
\alpha=\frac{p+2}{2-p}\, ,
\end{equation}
which is clearly negative for $p>2$ (i.e. $\epsilon<0$).  

Before proceeding, let us make a comment on the above reconstruction procedure. Actually, 
this procedure is based on the Hamilton-Jacobi-like approach independently introduced in  
\cite{Muslimov:1990be,Salopek:1990jq}, however now the difference is that instead of 
solving the resulting first-order nonlinear differential equation for $H(\phi)$ (or 
equivalently $H(t)$) we impose it at will, and we calculate the potential 
$V(\phi)$ required for the given dynamical evolution. Strictly speaking, one should 
examine whether there are other solutions corresponding to the same potential, but for 
the purposes of this work it is adequate to extract one of them.

We now calculate the effective equation of state 
\begin{equation}
\label{eos}
w_\mathrm{eff}=\frac{p}{\rho}=-1-\frac{2\dot{H}}{3H^2} \, .
\end{equation}
Using the Hubble parameter (\ref{IV1}) we obtain 
\begin{equation}
\label{eos-1}
w_\mathrm{eff}=-1+\frac{2 b_0 (-t+t_s)^{-1+\alpha } \alpha }{3 \left[c_0+b_0 
(-t+t_s)^{\alpha }\right]^2}\, .
\end{equation}
In one of the interesting cases at hand, namely when $\alpha\gg 1$, which corresponds to 
the Type IV singularity, the effective equation-of-state parameter becomes exactly equal 
to $w_\mathrm{eff}=-1$ at the time $t=t_s$. This case is problematic for the following 
reason: As one can see, the main source of the problem originates from the fact that 
according to (\ref{IV1}) we obtain $\dot{H}=0$ at $t=t_s$. This can potentially introduce 
a very large instability when crossing the phantom divide $w_\mathrm{eff}=-1$. In order 
to show this explicitly we introduce the variables $X_{\phi}$ and $Y$, defined as
\begin{equation}
\label{newvaribles}
X_{\phi}=\dot{\phi}\, , \quad Y=\frac{f(\phi)}{H}\, .
\end{equation}
In practice, the variable $Y$ quantifies the deviation from the solution of the 
reconstructed scalar-tensor theory, given in (\ref{ma11}). Using these new variables, 
the Friedmann equations (\ref{ma9}) and the field equation (\ref{ma12}) can be 
rewritten as
\be
\frac{\mathrm{d}X_{\phi}}{\mathrm{d}N}=\frac{f''(\phi)\left (X_{\phi}^2-1\right 
)}{2f'(\phi)H}-3\left (X_{\phi}-Y\right )\, , \quad 
\frac{\mathrm{d}Y}{\mathrm{d}N}=\frac{f'(\phi)\left(1-X_{\phi}Y\right)X_{\phi}}{H^2}\, ,
\label{newfrweqns}
\ee
where $N$ is the $e$-folding number. Hence, the scalar-tensor reconstruction solution of 
(\ref{ma11}) corresponds to the following values for the new variables:
\begin{equation}
\label{standardvalues}
X_{\phi}=1\, ,\quad Y=1\, ,
\end{equation}
which correspond exactly to the basic critical point of system (\ref{newfrweqns}). As 
usual, in order to examine when this solution is stable we linearly perturb it around 
this critical point as \cite{Copeland:1997et,Chen:2008ft,Leon2011}
\begin{equation}
\label{linearpert}
X_{\phi}=1+\delta X_{\phi}\, , \quad Y=1+\delta Y\, ,
\end{equation}
and thus the dynamical system (\ref{newfrweqns}) can be re-written 
as
\begin{equation}
\label{dynamicalsystem}
\frac{\mathrm{d}}{\mathrm{d}N}\left(
\begin{array}{c}
  \delta X_{\phi} \\
  \delta Y \\
\end{array}
\right)=\left(
\begin{array}{cc}
 -\frac{\ddot{H}}{\dot{H}H}-3 & 3
 \\ -\frac{\dot{H}}{H^2} & -\frac{\dot{H}}{H^2} \\
\end{array} \right)\left(
\begin{array}{c}
  \delta X_{\phi} \\
  \delta Y \\
\end{array}
\right)\, .
\end{equation}
Hence, the stability of the dynamical system (\ref{dynamicalsystem}) is ensured if the 
eigenvalues of the involved $2\times2$ matrix are negative. These eigenvalues in general 
read as
\begin{align}
\label{eigenvalues}
M_+=&\frac{1}{2} \left[-\left(\frac{H''(t)}{H'(t) 
H(t)}+\frac{H'(t)}{H(t)^2}+3\right) \right. \nn 
& \left. +\sqrt{\left(\frac{H''(t)}{H'(t) 
H(t)}+\frac{H'(t)}{H(t)^2}+3\right)^2-\frac{4 H''(t)}{H(t)^3}-\frac{12 H'(t)}{H(
t)^2}}\right] \, , \nn
M_{-}=&\frac{1}{2} \left[-\left(\frac{H''(t)}{H'(t) 
H(t)}+\frac{H'(t)}{H(t)^2}+3\right) \right. \nn 
& \left. -\sqrt{\left(\frac{H''(t)}{H'(t) 
H(t)}+\frac{H'(t)}{H(t)^2}+3\right)^2-\frac{4 H''(t)}{H(t)^3}-\frac{12 H'(t)}{H(
t)^2}}\right] \, ,
\end{align}
and thus for the Hubble rate  (\ref{IV1}), at the time $t\simeq t_s$, they become
\begin{equation}
\label{eigenvalues1}
M_+=0\, , \quad M_{-}=\frac{1}{2} \left[-6+\frac{2 (-1+\alpha )}{(-t+t_s)c_0}\right]\, .
\end{equation}
Since $M_->0$ when $t<t_s$ and $M_-<0$ when $t>t_s$, the dynamical system is unstable at 
the transition point $t=t_s$, which corresponds to the phantom-divide transition point. 
This result is valid for any value of the parameter $\alpha$. In addition, at the 
transition point the eigenvalue $M_-$ is infinite. This instability was first 
observed in \cite{vikman,Xia:2007km}. In order to evade this kind of problems in 
scalar-tensor cosmologies, the reconstruction method must be enriched with the 
presence of two scalars. As we show in the next subsection, the instability does not 
occur in such a case.

\subsection{Singular evolution with hilltop-like potentials: Description with two scalar 
fields}

As we demonstrated in the previous subsection, when we consider the transition from the 
non-phantom phase to the phantom one, in the context of scalar-tensor gravity with one 
scalar field, there appears an infinite instability at the transition point 
\cite{Nojiri:2005pu,Capozziello:2005tf,vikman,Xia:2007km}. However, this instability 
can easily be removed by considering scalar-tensor theories with two scalar fields.

Consider the following two-field scalar-tensor gravity, with action: 
\be
\label{A1}
S=\int d^4 x \sqrt{-g}\left\{\frac{1}{2\kappa^2}R
 - \frac{1}{2}\omega(\phi)\partial_\mu \phi \partial^\mu \phi
 - \frac{1}{2}\eta(\chi)\partial_\mu \chi\partial^\mu \chi
 - V(\phi,\chi)\right\}\, .
\ee
As in the single scalar case, the function $\omega(\phi)$ is the kinetic function for the 
scalar field $\phi$, and accordingly $\eta(\chi)$ is the kinetic function for the scalar 
field $\chi$. When the functions $\omega(\phi)$ or $\eta(\chi)$ are negative the 
corresponding scalar field becomes a ghost field. However, since such a field would 
lead to inconsistencies at the quantum level \cite{Cline:2003gs}, one expects it to arise 
through an effective description of a non-phantom fundamental theory \cite{Nojiri:2003vn}.

Assuming that the two scalar fields $\phi$ and $\chi$ depend only on the cosmic-time 
coordinate $t$, and also that the spacetime is described by a flat FRW metric of the form 
(\ref{JGRG14}), we easily write the Friedmann equations in the form
\begin{eqnarray}
\label{A2}
&&\omega(\phi) {\dot \phi}^2 + \eta(\chi) {\dot \chi}^2
= - \frac{2}{\kappa^2}\dot H \, , \\
&&
V(\phi,\chi)=\frac{1}{\kappa^2}\left(3H^2 + \dot H\right)\, .
\label{A2bb}
\end{eqnarray}
Following the procedure of the previous subsection, and generalizing it in the case of 
two fields, we impose the parametrization 
\be
\label{A3}
\omega(t) + \eta(t)=- \frac{2}{\kappa^2}f'(t)\, , \quad 
V(t)=\frac{1}{\kappa^2}\left[3f(t)^2 + f'(t)\right]\, ,
\ee
which is consistent with an explicit solution for (\ref{A2}) and (\ref{A2bb}) of the from 
\be
\label{A4}
\phi=\chi=t\, ,\quad H=f(t)\, .
\ee
For a detailed account on this reconstruction method see \cite{Nojiri:2005pu}. 

The kinetic functions $\omega (\phi)$ and $\eta (\chi)$ can be chosen in such a way 
that $\omega (\phi)$ is always positive and $\eta(\chi)$ is always negative, hence one 
field is canonical and the other one is ghost. This is exactly the realization of the 
quintom scenario \cite{Cai:2009zp}. In particular, a convenient choice for the kinetic 
functions is  
\begin{align}
\omega(\phi) =&-\frac{2}{\kappa^2}\left\{f'(\phi)
- \sqrt{\alpha_1(\phi)^2 + f'(\phi)^2} \right\}>0 \, , \nn 
\eta(\chi) =& -\frac{2}{\kappa^2}\sqrt{\alpha_1(\chi)^2 + f'(\chi)^2}<0\, ,
\label{A5}
\end{align}
where  $\alpha_1(x)$ is an arbitrary function of its argument. We now define a new 
function 
$\tilde f(\phi,\chi)$ through
\be
\label{A6}
\tilde f(\phi,\chi)\equiv - \frac{\kappa^2}{2}\left[\int d\phi 
\omega(\phi) + \int d\chi \eta(\chi)\right]\,,
\ee
which has the characteristic property
\be
\label{A7}
\tilde f(t)=f(t)\, ,
\ee
and therefore the constants of integration are fixed in (\ref{A6}). Assuming that 
$V(\phi,\chi)$ is given in terms of the function $\tilde f(\phi,\chi)$ as
\be
\label{A8}
V(\phi,\chi)=\frac{1}{\kappa^2}\left[3{\tilde f(\phi,\chi)}^2
+ \frac{\partial \tilde f(\phi,\chi)}{\partial \phi}
+ \frac{\partial \tilde f(\phi,\chi)}{\partial \chi} \right]\, ,
\ee
we find that, in addition to the Friedmann equations given in (\ref{A2}) and 
(\ref{A2bb}), 
the following two-scalar field equations are also satisfied:
\begin{eqnarray}
\label{A9}
&&0 = \omega(\phi)\ddot\phi + \frac{1}{2}\omega'(\phi) {\dot \phi}^2
+ 3H\omega(\phi)\dot\phi + \frac{\partial \tilde V(\phi,\chi)}{\partial 
\phi} \, , \\
&&
0 = \eta(\chi)\ddot\chi + \frac{1}{2}\eta'(\chi) {\dot \chi}^2
+ 3H\eta(\chi)\dot\chi + \frac{\partial \tilde V(\phi,\chi)}{\partial 
\chi}\, .
\label{A9bb}
\end{eqnarray}
In summary, upon using the above kinetic functions $\omega (\phi)$, $\eta (\chi)$ and the 
scalar potential $V(\phi,\chi)$, we have a two-scalar field scalar-tensor model with 
cosmological evolution of the form given in (\ref{A4}).
 
Let us now extend the analysis of the previous subsection, and proceed to the 
reconstruction of a two-field scalar-tensor gravity generating the cosmology described 
by the scale factor (\ref{IV1}). As we discussed, this class of cosmological evolution is 
related to a single-field scalar-tensor theory, which generates a hilltop-like 
scalar potential of the form (\ref{hilltop}) for the corresponding canonical scalar 
field. Moreover, it exhibits specific finite-time singularities, according to the 
values of $\alpha$ given in the list below (\ref{IV1}). 

In order to proceed, we need to impose an ansatz for the arbitrary function $\alpha_1(x)$ 
appeared in (\ref{A5}). We choose it as
\begin{equation}
\label{alphadef}
\alpha_1 (x)=\sqrt{b_0\alpha(-x +t_s)^{\alpha -1}-H'(x)^2}\, ,
\end{equation}
and therefore (\ref{A5}) give  
\begin{eqnarray}
&&\omega (\phi )=\frac{2 \left[b_0 \alpha  (t_s-\phi )^{-1+\alpha }+\sqrt{b_0 \alpha  
(t_s-\phi )^{-
1+\alpha }}\right]}{\kappa ^2}\, , \nonumber\\
&&
\eta (\chi)= -\frac{2 \sqrt{b_0 \alpha  (t_s-\chi )^{-1+\alpha }}}{\kappa ^2}\, .
\label{omeganforhilltopdouble}
\end{eqnarray}
In order to avoid inconsistencies we assume that $\alpha$ has the form
\be
\label{IV2}
\alpha= \frac{n}{2m + 1}\, ,
\ee
and therefore when $n$ is an odd integer, the kinetic function $\omega(\phi)$ is 
positive, while $\eta (\chi)$ is negative. The corresponding function $\tilde 
f(\phi,\chi)$ from (\ref{A6}) writes as
\begin{equation}
\label{barf}
\tilde f(\phi,\chi)=- b_0  (t_s-\phi )^{\alpha } 
-\frac{2 \sqrt{
b_0 \alpha  } (t_s-\phi )^{\frac{\alpha+1 }{2}} }{(\alpha +1) }
-\frac{2 \sqrt{
b_0 \alpha  } (t_s-\chi )^{\frac{\alpha+1 }{2}} }{(\alpha +1) }\, ,
\end{equation}
and thus the scalar potential $V(\phi,\chi)$ in (\ref{A8}) becomes
\begin{align}
\label{scalarpotentialhilltop}
V(\phi,\chi) & =\frac{b_0 \alpha    (t_s-\phi )^{\alpha-1} }{\kappa^2}  
+\frac{2 \sqrt{b_0 \alpha 
}(t_s-\phi 
)^{\frac{\alpha-1 }{2}}+\sqrt{b_0 \alpha } (\alpha-1 )\text{  }(t_s-\phi 
)^{\frac{\alpha-1}{2} 
} }{(1+\alpha )\kappa ^2}
\nn
& +   \frac{2 \sqrt{b_0 \alpha  } (t_s-\chi )^{\frac{\alpha-1}{2}}}{(1+\alpha ) \kappa 
^2}-\frac{  \sqrt{b_0 \alpha }(\alpha-1 ) (t_s-\chi 
)^{\alpha-1}}{(1+\alpha ) \kappa ^2} 
\nn
&+\frac{3}{\kappa^2}   \left\{ b_0  (t_s-\phi )^{\alpha } 
+\frac{2 \sqrt{
b_0 \alpha  } (t_s-\phi )^{\frac{\alpha+1 }{2}} }{(\alpha +1)  }
+\frac{2 \sqrt{
b_0 \alpha  } (t_s-\chi )^{\frac{\alpha+1 }{2}} }{(\alpha+1 )  }\right\}^2\, .
\end{align}

For the above two-field scalar-tensor theory, the solution (\ref{A4}) is stable as we 
will demonstrate in detail. Indeed, we introduce the quantities $X_{\phi}$, $X_{\chi}$ 
and 
$\bar{Y}$ through
\begin{equation}
\label{newvariables}
X_{\phi}=\dot{\phi}\, , \quad  X_{\chi}=\dot{\chi}\, , \quad  
\bar{Y}=\frac{\tilde{f}(\phi 
,\chi)}{
H}\, ,
\end{equation}
and thus the Friedmann equations (\ref{A2}), combined with (\ref{A9}) and (\ref{A9bb}), 
can be expressed as
\begin{align}
& \frac{\mathrm{d}X_{\phi}}{\mathrm{d}N}=\frac{\omega '(\phi)\left (X_{\phi}^2-1\right 
)}{2\omega(\phi)H}-3\left (X_{\phi}-\bar{Y}\right ) \, ,
\nn & 
\frac{\mathrm{d}X_{\chi}}{\mathrm{d}N}=\frac{\eta '(\chi)\left (X_{\chi}^2-1\right 
)}{2\eta(\chi)H}-3\left (X_{\chi}-\bar{Y}\right ) \, ,
\nn &\frac{\mathrm{d}\bar{Y}}{\mathrm{d}N}=\frac{3 X_{\phi} X_{\chi}\left 
(1-\bar{Y}^2\right 
)}{X_{\phi}+X_{\chi}}+\frac{\dot{H}}{H^2}\frac{X_{\phi}X_{\chi}+1-\bar{Y}(X_{\phi}+X_{\chi
})}{X_{ \phi}+X_{\chi}} \, .
\label{neweqs}
\end{align} 
The solution appearing in (\ref{A4}) corresponds to the following values of the 
variables $X_{\phi}$, $X_{\chi}$ and $\bar{Y}$:
\begin{equation}
\label{valuesofvariables}
X_{\phi}=1\, , \quad X_{\chi}=1\, , \quad \bar{Y}=1\, ,
\end{equation}
which correspond exactly to the basic critical point of system (\ref{neweqs}). As we did 
in the previous subsection, in order to examine when this solution is stable we 
linearly perturb it around this critical point as 
\begin{equation}
\label{dynamicalpertyrbationsii}
X_{\phi}=1+\delta X_{\phi} \, , \quad X_{\chi}=1+\delta X_{\chi}\, , \quad 
\bar{Y}=1+\delta \bar{Y}\, ,
\end{equation}
obtaining the perturbative dynamical system
\begin{equation}
\label{dynamicalsystemiii}
\frac{\mathrm{d}}{\mathrm{d}N}\left(
\begin{array}{c}
  \delta X_{\phi} \\
  \delta X_{\chi} \\
  \delta \bar{Y}\\
\end{array}
\right)=\left(
\begin{array}{ccc}
 -\frac{\omega' (\phi)}{H\omega (\phi)}-3 & 0 & 3
 \\ 0 &-\frac{\eta' (\chi)}{H\eta (\chi)} & 3 \\
 0 & 0 & -3-\frac{\dot{H}}{H^2} \\
\end{array} \right)\left(
\begin{array}{c}
   \delta X_{\phi} \\
  \delta X_{\chi} \\
  \delta \bar{Y}\\
\end{array}
\right) \, .
\end{equation}
The eigenvalues of the matrix appearing in the dynamical system 
(\ref{dynamicalsystemiii}) are found to be 
\begin{equation}
\label{eigenvaluesdoblescal}
M_{\phi}=-\frac{\omega' (\phi)}{H\omega (\phi)}-3\, , \quad {\,}M_{\chi}=-\frac{\eta' 
(\chi)}{H\eta 
(\chi)}-3\, , \quad M_{\bar{Y}}=-3-\frac{\dot{H}}{H^2} \, .
\end{equation}
Hence, in the case where the Hubble rate is given by (\ref{IV}), these eigenvalues 
become
\begin{align}
\label{eigendoublehill}
& M_{\phi}= -3-\frac{(-1+\alpha ) \left(1+2 \sqrt{b_0 \alpha  (t_s-\phi )^{-1+\alpha 
}}\right) \sqrt{
b_0 \alpha  (t_s-\phi )^{-1+\alpha }}}{2 c_0 \left[b_0 \alpha  (t_s-\phi )^{\alpha 
}+(t_s-\phi ) \sqrt{b_0 \alpha  (t_s-\phi )^{-1+\alpha }}\right]}
\, ,\nn &
M_{\chi}=-3-\frac{-1+\alpha }{2 c_0 (t_s-\chi )} \, ,\nn &
M_{\bar{Y}}=-3 \, ,
\end{align}
which are clearly negative for all values of the cosmic time $t$. Therefore, we deduce 
that the solution (\ref{A4}), with the Hubble rate given in (\ref{IV}), is stable. Hence, 
as we have already mentioned, the presence of two scalar fields is adequate to solve the 
instability problem of the one-field theory. Having the stable solution (\ref{A4}) at 
hand, we establish the fact that a Type IV singularity can underlie the cosmology 
generated by the Hubble rate (\ref{IV}). In particular, one possible description can be 
realized by a two-field generic scalar-tensor theory with kinetic functions given in 
(\ref{omeganforhilltopdouble}) and with the scalar potential appearing in 
(\ref{scalarpotentialhilltop}). 

Finally, it is worthy to mention that the solution (\ref{A4}) is only one particular 
solution of the dynamical system  (\ref{A2}) with (\ref{A9}) and (\ref{A9bb}). In 
principle, there might exist other solutions, however the stability of (\ref{A4}) which 
renders it an attractor, ensures that this solution will be the asymptotic limit of a 
large class of solutions. 

\subsection{Singular versus non-singular evolution}

Let us close this section by making some comments on the possible removal of the 
singularities discussed above. In the literature there are mechanisms that could remove 
some singularities of a  given cosmological scenario. Two of the known such mechanisms 
are 
the inclusion of quantum effects, or the suitable slight modification of the action (for 
instance changing the $F(R)$ or the potential form \cite{Damour:1997cb}). 

Indeed, singularities can become milder through the incorporation of quantum effects in a 
perturbative way, without changing the action. However this is not the case for the 
singularities analyzed in the previous subsections (in general even the quantum 
description of the phantom regime is ambiguous \cite{Cline:2003gs,Nojiri:2013ru}). 

Nevertheless, one could regularize and remove some of these singularities by modifying 
the action, i.e. by slightly changing the potential \cite{Damour:1997cb}. However, 
following this approach is not the scope of our analysis. In particular, we would like to 
mention that in principle the potential of a model is expected in general to be 
determined 
by the fundamental theory itself, thus in general one cannot modify it at will in a 
suitable way that will make the singularities to be absent. Hence, since any potential 
could arise from a fundamental theory, even if it is of special form, including the ones 
considered in this work, it is interesting and necessary to investigate their 
cosmological 
implications and in particular the realization of singularities, without entering the 
discussion to modify it, i.e modifying the theory. Indeed, a theory admitting weak 
singularities has to be special in some respect (for instance note that there are not 
such singularities in the vacuum general relativity case), however this could still be 
the case in Nature, deserving further investigation.

Additionally, note that there could still be cases where weak singularities cannot be 
removed even after a slight modification of the action. For instance, as it was shown 
in \cite{Appleby:2009uf}, the singularities arising in $F(R)$ gravity in the case where 
$F''(R_0)=0$ for some $R=R_0$, may not be removed through the procedure 
of \cite{Damour:1997cb} if $F'''(R_0)\neq0$. 

In summary, although removing the singularities is definitely an option, and a 
non-singular evolution can be realized, since singular evolution is a possibility 
too in this work we are interested to investigate the resulting cosmological behavior. 
Moreover, we desire to show that in some cases a singular evolution can be consistent 
and in agreement with the standard observed Universe history.

\section{Two-field scalar-tensor gravity: specific examples}
\label{variousexamples}

The use of two scalar fields offers a great number of possibilities for cosmological 
evolution, with various diverse features that were absent in the case of a single scalar 
field. In this section we are interested in reconstructing specific two-field models that 
may still exhibit some singularity types, in particular Type II and Type IV.

\subsection{Generic examples of two-field scalar-tensor theories exhibiting Type II
or Type IV singularities}
\label{subgenexamples}

Let us start by considering the Hubble rate to have the very general form 
\be
\label{IV1anoj}
H(t) = f_1(t) + f_2(t) \left( t_s - t \right)^\alpha\, .
\ee
We can choose the arbitrary functions $f_1(t)$ and $f_2(t)$ to be smooth and 
differentiable 
functions of $t$. In the following, and without loss of generality, we shall consider the 
case where $\alpha$ is given by two integers $n$ and $m$, namely
\be
\label{IV2noj}
\alpha= \frac{n}{2m + 1}\, ,
\ee
and specify furthermore the value of the integer $n$ when this is needed. For more 
general values 
of the parameter $\alpha$ we may extend (\ref{IV1anoj}) to
\be
\label{IV1Bnoj}
H(t) = f_1(t) + f_2(t) \left| t_s - t \right|^\alpha\, .
\ee
 
Concerning the kinetic functions $\omega (\phi)$ and $\eta (\phi)$ of the two fields that 
appear in action (\ref{ma7}), we use
\be
\omega(\phi) =-\frac{2}{\kappa^2}f_1'(\phi) \, , \quad 
\eta(\chi) = -\frac{2}{\kappa^2} \left[ f_2'(\chi) \left( t_s - \chi \right)^\alpha
+ \alpha f_2(\chi) \left( t_s - \chi \right)^{\alpha - 1} \right]\, , 
\label{B1noj}
\ee
which implies that the function $\alpha_1(x)$ appearing in (\ref{A5}) is of the form
\begin{equation}
\label{alphaxnewdef}
\alpha_1(x)=\sqrt{\left [ f_2 (x)\left (t_s-x \right )^{\alpha}+\alpha f_2(x)\left 
(t_s-x\right )^{\alpha-1}\right ]^2-\left[ H'(x)\right]^2}\, .
\end{equation}
Here, the variable $x$ can be either the scalar field $\phi$ or $\chi$. Having chosen 
$\alpha_1(x)$ as in expression (\ref{alphaxnewdef}) allows to specify the final form of 
the function $\tilde{f}(\phi,\chi)$, defined in  (\ref{A6}), and indeed in the case at 
hand becomes
\be
\label{B2noj}
\tilde f(\phi,\chi) = f_1(\phi) + f_2(\chi) \left( t_s - \chi \right)^\alpha\, .
\ee
The corresponding scalar potential of (\ref{A8}) reads
\be
\label{B3noj}
V(\phi,\chi)=\frac{1}{\kappa^2}\left\{3\left[
f_1(\phi) + f_2(\chi) \left( t_s - \chi \right)^\alpha \right]^2 
+ f_1'(\phi) +  f_2'(\chi) \left( t_s - \chi \right)^\alpha
+ \alpha f_2(\chi) \left( t_s - \chi \right)^{\alpha - 1} \right\}\, .
\ee
Additionally, the kinetic functions from (\ref{B1noj}) now become
\begin{align}
\label{kineticfunct}
& \omega (\phi)= -\frac{2 f_1'(t)}{\kappa ^2},\nn &
\eta (\chi)= -\frac{2 \left[\alpha  (t_s-\chi )^{-1+\alpha } f_2(\chi )+(t_s-\chi 
)^{\alpha } f_2'(
t)\right]}{\kappa ^2} \, .
\end{align}
Finally, from the form of the Hubble rate (\ref{IV1Bnoj}), we can straightforwardly 
extract the total equation-of-state parameter of the universe  (\ref{eos}) as 
\begin{equation}
\label{eosnoj1}
w_{\mathrm{eff}}=-1-\frac{2 \left[f_1'(t)+(-t+t_s)^{\alpha } 
f_2'(t)   -(-t+t_s)^{-1+\alpha } \alpha  
f_2(t)\right]}{3 \left[f_1(t)+(-t+t_s)^{\alpha } f_2(t)\right]{}^2}\, .
\end{equation}

As we already mentioned, the functions $f_1(t)$ and $f_2(t)$ are arbitrary functions, so 
a convenient choice might lead to interesting results, regarding the cosmological 
evolution. Assuming that the Hubble rate is equal to $H(t)=f_1(t)+f_2(t) $, 
one convenient choice for $f_1(t)$ and $f_2(t)$ could be
\begin{equation}
\label{newchocie1}
f_1(t)=f^\phi (\phi)\, , \quad f_2(t)=f^\chi (\chi) \, ,
\end{equation}
where
\be
\label{B4noj}
f^\phi (\phi) = f_1^\phi (\phi) + f_2^\phi (\phi) \left( t_1 - \phi \right)^\alpha \, , 
\quad 
f^\chi (\chi) = f_1^\chi (\chi) + f_2^\chi (\chi) \left( t_2 - \chi \right)^\beta \, , 
\ee
and where $f_1^\phi (\phi)$, $f_2^\phi (\phi)$, $f_1^\chi (\chi)$, and $f_2^\chi (\chi)$ 
are smooth functions of the scalar fields. Then, in terms of the functions appearing in 
(\ref{B4noj}), the Hubble rate is written as  
\be
\label{B5noj}
H(t) = f^\phi (t) + f^\chi (t) = f_1^\phi (t) + f_1^\chi (t) + f_2^\phi (t) \left( t_1 - 
\phi \right)^\alpha + f_2^\chi (t) \left( t_2 - \phi \right)^\beta \, .
\ee

An interesting cosmological scenario can be realized if we choose $t_1 \ll t_2$, 
$\alpha>1$ and $\beta <0$. In this case, $t_1$ could correspond to a time around the 
inflation era, while $t_2$ to the dark energy epoch, present or future. In 
addition, when $\alpha>1$ and $\beta <0$ it easily follows from the Hubble rate 
(\ref{B5noj}) that at $t_1$ the physical system of the two scalars experiences a Type IV 
singularity, while at $t_2$, it experiences a Type I (Big Rip) singularity, if 
$\beta<-1$, 
or a Type III, if $-1<\beta<0$. It is worthy to further specify the functions appearing 
in 
(\ref{B4noj}) and quantitatively examine the cosmological evolution and the resulting 
physical picture. As an example we choose
\be
\label{B6noj}
f^\phi (\phi) = \frac{f_1}{\sqrt{t_0^2 + \phi^2}} + \frac{f_2 \phi^2}{t_0^4 + 
\phi^4}\left( t_1 - \phi \right)^\alpha\, , \quad 
f^\chi (\chi) = f_3 \left( t_2 - \chi \right)^\beta \, , 
\ee
where $t_0$, $f_1$, $f_2$, and $f_3$ are considered to be positive constants, so that 
$H(t)>0$. For the choice (\ref{B6noj}), the Hubble rate becomes
\begin{equation}\label{hubnoj1}
H(t)=\frac{f_1}{\sqrt{t^2+t_0^2}}+\frac{f_2 t^2 (-t+t_1)^{\alpha }}{t^4+t_0^4}+f_3 
(-t+t_2)^{\beta }
 \, ,
\end{equation}
and the corresponding kinetic functions of the two scalars read
\begin{align}
\label{kineticfs}
& \omega (\phi)=-\frac{2 }{\kappa ^2}\left[-\frac{f_1 
t}{\left(t^2+t_0^2\right)^{3/2}}-\frac{4 f_2 
t^5 (-t+t_1)^{\alpha }}{\left(t^4+t_0^4\right)^2}+\frac{2 f_2 t (-t+t_1)^{\alpha 
}}{t^4+t_0^4}-\frac{f_2 t^2 (-t+t_1)^{-1+\alpha } \alpha }{t^4+t_0^4}\right] \, ,
\nn &
\eta (\chi )=\frac{2 f_3 (-t+t_2)^{-1+\beta } \beta }{\kappa ^2} \, .
\end{align}
Since the parameter $\beta$ is assumed to be $\beta<0$, the kinetic function $\eta 
(\chi)$ is negative, and therefore $\chi$ is a ghost field. On the contrary, 
$\omega(\phi)$ is positive and therefore $\phi$ is a canonical scalar field, as long as 
$f_2\ll f_1$. The corresponding $\tilde{f}(\phi,\chi)$ function from (\ref{B2noj}) writes 
as
\begin{equation}
\label{tildefnewfunction}
\tilde{f}(\phi,\chi)=\frac{f_1}{\sqrt{t_0^2+\phi ^2}}+\frac{f_2 (t_1-\phi )^{\alpha } 
\phi 
^2}{t_
0^4+\phi ^4}+f_3 (t_2-\chi )^{\beta } \, ,
\end{equation}
and therefore the scalar potential (\ref{B3noj}) is given by
\begin{align}
\label{scalarpotentialnew}
V(\phi,\chi)=& \frac{1}{\kappa ^2}\left[-\frac{f_1 \phi }{\left(t_0^2+\phi 
^2\right)^{3/2}}-\frac{4 
f_2 (t_1-\phi )^{\alpha } \phi ^5}{\left(t_0^4+\phi ^4\right)^2}+\frac{2 f_2 (t_1-\phi 
)^{\alpha } \phi }{t_0^4+\phi ^4}-\frac{f_2 \alpha  (t_1-\phi )^{-1+\alpha } \phi 
^2}{t_0^4+\phi ^4}\right]
\nn
&
+\frac{3}{\kappa ^2} \left\{\left[\frac{f_1}{\sqrt{t_0^2+\phi ^2}}+\frac{f_2 (t_1-\phi 
)^{\alpha } \phi ^2}{t_0^4+\phi ^4}+f_3 (t_2-\chi )^{\beta }\right]^2-f_3 \beta  
(t_2-\chi 
)^{-1+\beta }\right\} \, .
\end{align}
Finally, the total equation-of-state parameter corresponding to the Hubble rate 
(\ref{hubnoj1}) reads
\begin{equation}
\label{weffnewexample}
w_{\mathrm{eff}}=-1-\frac{2 \left[-\frac{f_1 t}{\left(t^2+t_0^2\right)^{3/2}}-\frac{4 f_2 
t^5 (-t+t_
1)^{\alpha }}{\left(t^4+t_0^4\right)^2}+\frac{2 f_2 t (-t+t_1)^{\alpha 
}}{t^4+t_0^4}-\frac{f_2 t^2 (
-t+t_1)^{-1+\alpha } \alpha }{t^4+t_0^4}-f_3 (-t+t_2)^{-1+\beta } \beta \right]}{3 
\left[\frac{f_1}{
\sqrt{t^2+t_0^2}}+\frac{f_2 t^2 (-t+t_1)^{\alpha }}{t^4+t_0^4}+f_3 (-t+t_2)^{\beta 
}\right]^2} \, .
\end{equation}

In the early universe $f^\phi(t)$ dominates in $H(t)$ in (\ref{B5noj}), and therefore the 
inflationary era occurs, with a Type IV singularity at $t=t_1$.  In the late-time 
universe $f^\chi(t)$ dominates in $H(t)$ in (\ref{B5noj}), and the universe will end with 
a Type I (Big Rip) singularity ($\beta<-1$) or Type III singularity ($-1<\beta<0$) at 
$t=t_2$. Having made these assumptions, it is worthy to study the behavior of the EoS 
(\ref{weffnewexample}) as a function of the cosmic time. Let us first consider time 
scales near the Type IV singularity, thus $t\sim t_1$. In such a 
case, by 
disregarding subdominant terms, the EoS takes the form
\begin{equation}
\label{eosinfex1}
w_{\mathrm{eff}}\simeq -1-\frac{2 \left[-\frac{f_1 t}{\left(t_0^2\right)^{3/2}}-f_3 
(t_2)^{-1+\beta 
} \beta \right]}{3 \left[\frac{f_1}{\sqrt{t_0^2}}+f_3 (t_2)^{\beta }\right]^2} \, ,
\end{equation}
and since $t_2\gg 1$ and $\beta<0$ it finally becomes
\begin{equation}
\label{quint1}
w_{\mathrm{eff}}\simeq -1+\frac{2 t}{3 f_1 t_0} \, ,
\end{equation}
which describes quintessential inflation. In the case where $t_0\gg 1$, the EoS 
(\ref{quint1}) becomes approximately $w_{\mathrm{eff}}\simeq -1$, i.e. it is effectively 
a cosmological constant. Furthermore, with regards to the late-time behavior, which 
corresponds to cosmic times near $t_2$, we shall assume that $t_2>t_0$ and also that 
$\alpha$ is of the form given in (\ref{IV2noj}), namely $\alpha= n/(2m + 1)$. 
For $n$ even, and $2<\alpha <3$, the effective equation of state reads
\begin{equation}
\label{effeqnofstate1}
w_{\mathrm{eff}}\simeq -1+\frac{2 f_3 (-t+t_2)^{-1+\beta } \beta }{3 \left[f_2 
t^{-2+\alpha }+f_3 (-t+t_2)^{\beta }\right]^2}\, ,
\end{equation}
which is $w_{\mathrm{eff}}<-1$ when $t<t_2$. Therefore, eventually this case describes 
phantom dark energy at a time before the singularity. Additionally, the case $\alpha >3$ 
and $n$ even is slightly more complicated. In this case we set $x=-t+t_2$, and the EoS 
becomes
\begin{equation}
\label{effeqnofstate12}
w_{\mathrm{eff}}\simeq -1-\frac{2 (t_2-x)^{1-\alpha } (-2+\alpha )}{3 f_2} \, ,
\end{equation}
which for $t<t_2$ (or $x>0$) is $w_{\mathrm{eff}}<-1$, which again describes phantom 
evolution. In the case where $n$ is an even integer and $\alpha <3$, the EoS reads
\begin{equation}
\label{n1eqoa}
w_{\mathrm{eff}}\simeq -1+\frac{2 f_3 (-t+t_2)^{-1+\beta } \beta }{3 \left[f_2 
t^{-2+\alpha }-f_3 (-
t+t_2)^{\beta }\right]^2},
\end{equation}
which describes phantom acceleration (recall that $\beta<0$), while for $\alpha>3$ we have
\begin{equation}
\label{eosnew123}
w_{\mathrm{eff}}\simeq -1-\frac{2 (t_2-x)^{1-\alpha } (2+\alpha )}{3 f_2}\, ,
\end{equation}
which describes phantom acceleration too. In summary, with two scalar fields we may 
realize a plethora of cosmological expansions, with finite-time singularities present in 
the cosmological evolution.

We close this subsection with another model with an interesting cosmological evolution. 
In particular, we consider the Hubble rate to be
\be
\label{B8noj}
H(t) = f_1^\phi(t) + f_1^\chi (t) + f_2(t) \left( t_s - t \right)^\alpha\, .
\ee
The function $f_1^\phi(t)$ can be chosen in such a way that it describes the (smooth) 
inflation era, and in addition the function $f_1^\chi(t)$ can be chosen to describe the 
present time accelerating expansion of the Universe. We can consider the following two 
classes of models, in which the functions are:
\begin{align}
\label{B9noj}
\mbox{(A):}\, &
f^\phi(\phi) = f_1^\phi(\phi) + f_2(\phi) \left( t_s - \phi \right)^\alpha\, , \quad 
f^\chi(\chi) = f_1^\chi (\chi) \, , \\
\label{B10noj}
\mbox{(B):}\, &
f^\phi(\phi) = f_1^\phi(\phi) \, , \quad 
f^\chi(\chi) = f_1^\chi (\chi) + f_2(\chi) \left( t_s - \chi \right)^\alpha \, .
\end{align}
Obviously, both models (A) and (B) generate identical evolution of the Universe's 
expansion, since the Hubble rate is the same. In model (A), the scalar field $\phi$ can 
be selected in order to generate both the inflationary era and the finite time 
singularity. In addition, in model (B) $\chi$ can generate both the present accelerating 
expansion, as well as the finite time singularity. We may choose $t_s$ to be large 
enough, in order to correspond to the far future, and moreover the function $f_2(t)$ can 
be chosen in order to give a small contribution to the present cosmological 
evolution of the Universe, and hence it is difficult for it's contribution to be 
observed. In the following, we shall present a model for which this scenario can be 
realized.

For brevity, let us consider model (\ref{B9noj}). The Hubble rate is simplified as
\begin{equation}
\label{finhu1}
H(t)=f_1^{\phi}(t)+(-t+t_1)^{\alpha } f_2^{\phi}(t)+f_1^{\chi}(t) \, ,
\end{equation}
and the kinetic functions of the scalar-tensor theory read
\begin{equation}
\label{kinetfin1}
\omega (\phi )=-\frac{2 \left[-(-t+t_1)^{-1+\alpha } \alpha  
f_2^{\phi}(t)+{f'}_1^{\phi}(t)+(-t+t_1)
^{\alpha } {f'}_2^{\phi}(t)\right]}{\kappa ^2} \, ,
\end{equation}
while
\begin{equation}
\label{kinetfin2}
\eta (\chi )=-\frac{2 {f'}_1^{\chi}(t)}{\kappa ^2} \, .
\end{equation}
The corresponding scalar potential (\ref{B3noj}) becomes
\begin{align}
\label{sclarpotent}
V(\phi,\chi)=&-\frac{\alpha  }{\kappa ^2}\left\{(t_1-\phi )^{-1+\alpha } f_2^{\phi}(\phi 
)+3 \left[f_
1^{\phi}(\phi )+(t_1-\phi )^{\alpha } f_2^{\phi}(\phi )+f_1^{\chi}(\chi 
)\right]^2\right.\nonumber\\
&\left. 
+{f'}_1^{\phi}(\phi )+(t_1-\phi )^{\alpha } 
{f'}_2^{\phi}(\phi 
)+{f'}_1^{\chi}(\chi )\right\} \, ,
\end{align}
while the total EoS writes as
\begin{equation}
\label{fineqnofstate}
w_{\mathrm{eff}}=-1-\frac{2 \left[-(-t+t_1)^{-1+\alpha } \alpha  
f_2^{\phi}(t)+{f'}_1^{\phi}(t)+(-
t+t_1)^{\alpha } {f'}_2^{\phi}(t)+{f'}_1^{\chi}(t)\right]}{3 
\left[f_1^{\phi}(t)+(-t+t_1)^{\alpha } 
f_2^{\phi}(t)+f_1^{\chi}(t)\right]^2} \, .
\end{equation}
A convenient choice for the functions appearing in the Hubble rate (\ref{finhu1}) is  
\begin{equation}
\label{specificchocies}
f_1^{\phi}(t)=\frac{f_1}{\sqrt{t^2+t_0^2}}\, , \quad f_2^{\phi}(t)= \e^{-t \beta }\, , 
\quad f_1^{\chi}(t)=\frac{f_2 t^2}{t^4+t_0^4}\, ,
\end{equation}
and in this case we finally acquire
\begin{equation}
\label{eostatefin}
w_{\mathrm{eff}}=-1-\frac{2 \left[-\frac{f_1 t}{\left(t^2+t_0^2\right)^{3/2}}-\frac{4 f_2 
t^5}{\left(t^4+t_0^4\right)^2}+\frac{2 f_2 t}{t^4+t_0^4}- \e^{-t \beta } 
(-t+t_1)^{-1+\alpha } \alpha - \e^{-t \beta } (-t+t_1)^{\alpha } \beta \right]}{3 
\left[\frac{f_1}{\sqrt{t^2+t_0^2}}+\frac{f_2 t^2}{
t^4+t_0^4}+ \e^{-t \beta } (-t+t_1)^{\alpha }\right]^2} \, .
\end{equation}
Thus, we can now study it's behavior, focusing mainly at times corresponding to the 
inflationary era, as well as at late times. 

We first consider that $t_1$ corresponds to the inflationary regime, and that at this 
point a Type IV singularity occurs. This implies that $\alpha>1$ and moreover we assume 
that $\beta >0$. Furthermore, if $t_0>t_1$ then the EoS at early times reads
\begin{equation}
\label{effealry}
w_{\mathrm{eff}}=-1+\frac{2 t (-2 f_2+f_1 t_0)}{3 f_1^2 t_0^2} \, .
\end{equation}
In the case where $f_1 t_0>2f_2$, the Universe accelerates in a quintessential way near 
the Type IV singularity at $t_1$. In addition, if $t_0\gg 1$ then the second term in 
(\ref{effealry}) is negligible, and therefore $w_{\mathrm{eff}}\simeq -1$, which 
corresponds to a de Sitter expansion. Definitely, if $f_1 t_0<2f_2$ then the EoS 
(\ref{effealry}) describes phantom inflation.

Let us now focus at late times, where the effective EoS behaves as 
\begin{equation}
\label{eosfinalrelatlate}
w_{\mathrm{eff}}\simeq -1+\frac{2 t}{3 f_1 \sqrt{t^2+t_0^2}} \, ,
\end{equation}
which describes quintessential acceleration. In the case where $t\simeq t_0$, and $t_0\gg 
1$, relation (\ref{eosfinalrelatlate}) becomes a nearly de Sitter late-time acceleration, 
with $w_{\mathrm{eff}}\simeq -1$. It is worthy to say that the scalar field $\phi$ is 
the one with the positive kinetic term in all cases, with $\omega (\phi)$ being equal to
\begin{equation}
\label{omegaphikanonic}
\omega (\phi)= -\frac{2 \left[-\frac{f_1 t}{\left(t^2+t_0^2\right)^{3/2}}- \e^{-t \beta } 
(-t+t_1)^{
-1+\alpha } \alpha - \e^{-t \beta } (-t+t_1)^{\alpha } \beta \right]}{\kappa ^2} \, ,
\end{equation}
while $\eta (\chi)$ reads
\begin{equation}
\label{etachiscalarfield}
\eta (\chi)=\frac{4 f_2 \left(t^5-t t_0^4\right)}{\left(t^4+t_0^4\right)^2 \kappa ^2} \, ,
\end{equation}
and since $t_0\gg 1$ it can be negative for a wide range of $t$ values (we assume that $f_
2>0$). Therefore, the use of multiple scalar fields can significantly increase the 
cosmological scenarios for the physical system described by a specific Hubble rate.

\subsection{Slow-roll with two-field scalar-tensor theories}
\label{subslowroll}

In this subsection we qualitatively describe how the slow-roll approximation is affected 
by the use of two scalar fields describing the cosmological evolution. In addition, we 
discuss how the singularities affect the general slow-roll evolution and also we 
investigate when these can lead to inconsistencies. In principle, the results can be 
model-dependent, but using some illustrative examples we shall see how a singularity can 
affect the inflation parameters. Our study will remain at a qualitative level, since a 
detailed investigation would require the analysis of two- and three-
point statistics of field perturbations, which can be related to the two- and three- 
point 
statistics of the curvature perturbations on uniform density hypersurfaces. This could be 
achieved by using the $\delta$N-formalism 
\cite{Starobinsky:1982ee,Starobinsky:1982ee1,KofmanStarobinsky,Lyth:2005fi,
Vernizzi:2006ve,Choi:2007su,
Alabidi:2006qa,Alabidi:2005qi,
Sasaki:1998ug}, however this detailed study lies beyond the illustrative scopes of this 
article. For details the reader is referred to 
\cite{Starobinsky:1982ee,Starobinsky:1982ee1,KofmanStarobinsky,Lyth:2005fi,
Vernizzi:2006ve,Choi:2007su,
Alabidi:2006qa,
Alabidi:2005qi,Sasaki:1998ug}.

The most important inconsistency that the singularities might cause, is traced in the 
definition of the $e$-folding number $N$. The inconsistency occurs when the singularity 
appears during the inflationary era, even in the case the singularity is of Type IV. 
Therefore, we shall assume that the singularity occurs at exactly the time when inflation 
ends. In the end of this subsection we shall discuss in detail this problematic issue.

In general, single-field and multi-field inflation are very different, when the 
corresponding scenarios are viewed as dynamical systems. In the single-field case, the 
phase space formed by the slow-roll solution is one-dimensional, implying that the 
slow-roll solution is the unique attractor of the dynamical system. On the other hand, 
in the case where two-field inflation is studied, the phase space is two dimensional and 
the stable trajectories form an infinite countable space 
\cite{Starobinsky:1982ee,Starobinsky:1982ee1,KofmanStarobinsky,Lyth:2005fi,
Vernizzi:2006ve,Choi:2007su,
Alabidi:2006qa,
Alabidi:2005qi,Sasaki:1998ug}. Hence, the field values at the end of inflation will 
generally depend on the choice of the classical field trajectory. The corresponding 
$e$-folding number will determine the evolution along one of these classical trajectories 
\cite{Starobinsky:1982ee,Starobinsky:1982ee1,KofmanStarobinsky,Lyth:2005fi,
Vernizzi:2006ve,Choi:2007su,
Alabidi:2006qa,Alabidi:2005qi,Sasaki:1998ug}. In order to see this in a qualitative way, 
in this subsection we shall investigate when the slow-roll condition is violated in the 
cases of single-field and two-field inflation. The violation of the slow-roll condition 
will actually determine when inflation ends. We quantify our results by studying the 
behavior of the slow-roll inflation parameter $\epsilon$. 

\begin{figure}[h]
\centering
\includegraphics[width=22pc]{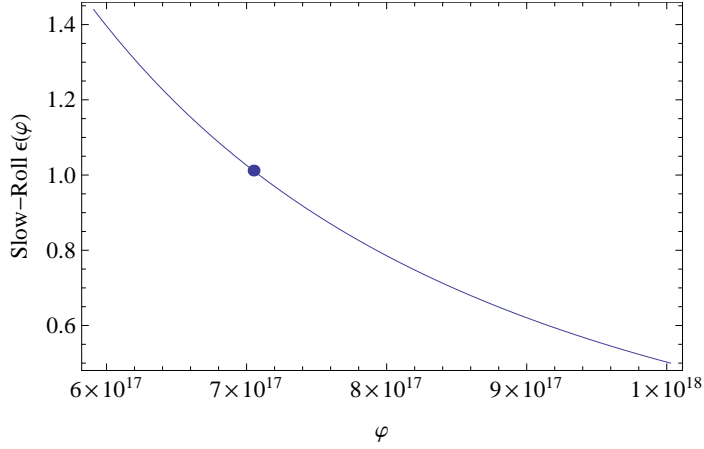}
\caption{{\it{The slow-roll parameter $\epsilon (\varphi)$ as a function of the scalar 
field $\varphi$, for the potential (\ref{largefieldpotent}) for $n\sim 1.95$. As the 
scalar field decreases the cosmological time is assumed to increase. The dot corresponds 
to the critical field value $\varphi_{e}=t_1+ \frac{ n}{ \sqrt{2} \kappa }$, at 
which the slow-roll condition is violated.}}}
\label{plot1}
\end{figure}

As a toy single-field example we shall consider a viable inflationary model, which 
belongs to the class of large-field inflationary realizations. In a previous article we 
performed a thorough analysis on this type of scenarios \cite{Nojiri:2015fra}, and thus 
we 
now focus on the qualitative analysis of the slow-roll parameter $\epsilon$. The model 
consists of a canonical scalar field $\varphi $, and the potential for large field values 
is given by
\begin{equation}
\label{largefieldpotent}
V(\varphi )\simeq d (\varphi-t_1)^n,
\end{equation}
with $t_1$ the time at which the Type IV singularity occurs, which for consistency is 
chosen to coincide with the time that inflation ends. As was demonstrated in 
\cite{Nojiri:2015fra} the Type IV singularity can be consistently accommodated in the 
cosmological evolution of the model under study. The slow-roll parameter $\epsilon$ in 
the 
case of a single canonical scalar field reads as
\begin{equation}
\label{singlescalar}
\epsilon=\frac{1}{2\kappa^2}\left[\frac{V'(\varphi)}{V(\varphi)}\right],
\end{equation}
and thus for the potential (\ref{largefieldpotent}) it becomes
\begin{equation}
\label{epsilonsingle}
\epsilon(\varphi)=\frac{n^2}{2 \kappa^2 (t_s-\varphi )^2}\, .
\end{equation}
It is easy to analytically calculate when inflation ends, which happens at a field 
value $\varphi_{e}$ for which $\epsilon (\varphi_{e})\sim 1$. Indeed, the slow-roll 
condition is violated when the scalar field exceeds the critical value $\varphi_{e}$, 
namely
\begin{equation}
\label{varphieccrit}
\varphi_{e}-t_1= \frac{ n}{\sqrt{2} \kappa }\, .
\end{equation}
In order to acquire an overall picture of the phenomenon, in Fig. \ref{plot1} we depict 
the behavior of $\epsilon (\varphi)$ as a function of $\varphi $. Notice that the large 
field values correspond to early times, and as the field value decreases, the 
cosmological time increases. Hence, the $x$-axis should be viewed backwards, as time 
increases.
As we can see, the slow-roll condition is violated at the field value $\varphi_{e}=t_1+ 
\frac{ n}{ \sqrt{2} \kappa }$, which is indicated by a dot on the graph. Thus, in 
the single-scalar case the ending of slow roll is characterized by a single point in the 
corresponding phase space. This fact has a direct impact on the slow-roll parameter 
$\epsilon (\varphi)$, as can be seen in Fig. \ref{plot1}.

Let us now come to the two-field case, which as we show is more involved. In 
order to simplify things we choose a simple example, and we consider the Hubble rate to be
\begin{equation}
\label{hubtwonew}
H(t)=f_1(t)+f_2(t)\, ,
\end{equation}
with $f_1(t)$ and $f_2(t)$ being equal to 
\begin{equation}
\label{deffunctions}
f_1(t)=\nu_1\left(-t+t_1\right)^{\alpha},{\,}{\,}{\,}f_2(t)=\nu_2\left(-t+t_1\right)^{
\alpha },
\end{equation}
where $\nu_1$ and $\nu_2$ are parameters. In this case the Type IV singularity occurs if 
$\alpha>1$. Thus, we can accommodate the Type IV singularity in a generalized two-field 
scalar-tensor theory with action given in (\ref{A1}), by following the reconstruction 
method of the previous section, with $\alpha_1(x)$ chosen as in (\ref{alphaxnewdef}). In 
particular, the kinetic functions $\omega (\phi)$ and $\eta (\chi)$ are equal to
\be
\label{B1nojkineticterms}
\omega(\phi) =-\frac{2}{\kappa^2}f_1'(\phi)\, ,\quad
\eta(\chi) = -\frac{2}{\kappa^2}  f_2'(\chi) \, , 
\ee
and by substituting the functions $f_{1,2}(t)$ from  (\ref{deffunctions}) we acquire
\be
\label{B1nojkineticterms1}
\omega(\phi) =\frac{2 \nu_1 (-t+t_1)^{\alpha-1} \alpha }{\kappa ^2}\, ,\quad
\eta(\chi) = \frac{2 \nu_2 (-t+t_1)^{\alpha-1 } \alpha }{\kappa ^2}\, .
\ee
Therefore, the corresponding scalar potential $V(\phi,\chi)$ writes as
\begin{align}
\label{scalarpotfnt}
V(\phi,\chi) = & -\frac{\nu_1 \alpha  (t_1-\phi )^{-1+\alpha }}{\kappa ^2}+\frac{3 
\nu_1^2 
(t_1-\phi )^{2 \alpha }}{\kappa ^2}-\frac{\nu_2 \alpha  (t_1-\chi )^{-1+\alpha }}{\kappa 
^2} \nonumber\\
& +\frac{6 \nu_1 \nu_2 (t_1-\phi )^{\alpha } (t_1-\chi )^{\alpha 
}}{\kappa ^2}+\frac{3 \nu_2^2 (t_1-\chi )^{2 \alpha 
}}{\kappa ^2}\, .
\end{align}
We assume that $t_1$ is identical with the cosmological time when inflation ends, and 
thus it is a very small number. In addition, we assume that inflation occurs for large 
field values, and thus at the time when inflation occurs, $t_1-\phi \gg 1$ and 
$t_1-\chi \gg 1$. Since $a>1$, the potential can be simplified to a great extent by 
keeping the dominant terms, becoming
\begin{align}
\label{scalarpotfnt1_0}
& V(\phi,\chi)=\frac{3 \nu_1^2 (t_1-\phi )^{2 \alpha }}{\kappa ^2} +\frac{6 \nu_1 \nu_2 
(t_1-\phi )^{\alpha } (t_1-\chi )^{\alpha }}{\kappa ^2}+\frac{3 \nu_2^2 (t_1-\chi )^{2 
\alpha }}{\kappa ^2}\, .
\end{align}

As a next step, we can transform the above two-field scalar-tensor theory to one that 
contains only canonical scalar fields. Indeed, by making the transformation
\begin{equation}
\label{transnewsec1}
t_1-\sigma=\int^{\chi}\sqrt{\eta(\chi)}\mathrm{d}\chi\, , \quad t_1-\varphi=\int^{\phi}
\sqrt{\omega(\phi)}\mathrm{d}\phi\, ,
\end{equation}
we can extract the identifications 
\begin{equation}
\label{identne1}
-\phi+t_1=\alpha_1\left(\varphi-t_1\right)^{\frac{2}{\alpha+1}},{\,}{\,}{\,}
-\chi+t_1=\alpha_2\left(
\sigma-t_1\right)^{\frac{2}{\alpha+1}}\, ,
\end{equation}
where $\alpha_{1}$ and $\alpha_2$ stand for 
\begin{equation}
\label{newscalarparame}
\alpha_1=\left[\frac{2\sqrt{2\alpha 
\nu_1}}{(1+\alpha)\kappa}\right]^{\frac{2}{\alpha+1}}\, , \quad 
\alpha_2=\left[\frac{2\sqrt{2\alpha 
\nu_2}}{(1+\alpha)\kappa}\right]^{\frac{2}{\alpha+1}}\, .
\end{equation}
Hence, in terms of the canonical scalar fields $\varphi$ and $\sigma$, the simplified 
scalar potential (\ref{scalarpotfnt1_0}) reads
\begin{align}
\label{scalarpotfnt1}
V(\varphi,\sigma)=&\frac{3 \nu_1^2 (\alpha_1)^{2\alpha}(\varphi-t_1 )^{\frac{4 
\alpha}{\alpha+1} }}{\kappa ^2} +\frac{6 \nu_1 \nu_2\left(\alpha_1 
\alpha_2\right)^{\alpha} (\varphi-t_1 )^{ \frac{2 
\alpha}{\alpha+1} } (\sigma-t_1 )^{\frac{2 \alpha}{\alpha+1} }}{\kappa 
^2}\nonumber\\
& +\frac{3 \nu_2^2 
(\alpha_2)^{2\alpha}(\sigma-t_1 )^{\frac{4 \alpha}{\alpha+1} }}{\kappa ^2}\, ,
\end{align}
while the canonical two-field action is simply
\begin{equation}\label{neactioncanscla}
S=\int d^4 x \sqrt{-g}\left\{\frac{1}{2\kappa^2}R
 - \frac{1}{2}\partial_\mu \varphi \partial^\mu \varphi
 - \frac{1}{2}\partial_\mu \sigma\partial^\mu \sigma
 - V(\varphi,\sigma)\right\}\, .
\end{equation}

\begin{figure}[h]
\centering
\includegraphics[width=20pc]{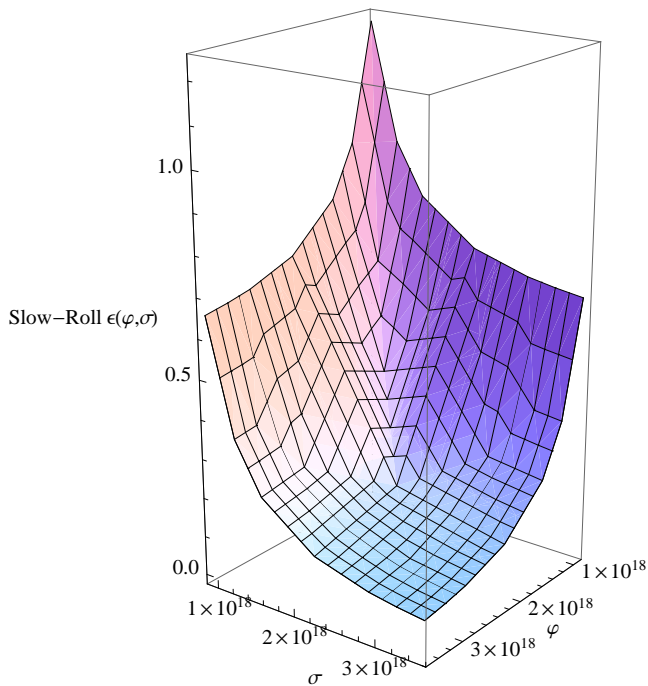}
\caption{\it{{The slow-roll parameter $\epsilon (\varphi,\sigma)$ as a function 
of the scalar fields $\varphi$ and $\sigma$, for the choice  
(\ref{deffunctions}) for $\nu_1=\nu_2$ and $\alpha\sim 1.1$. As the scalar field 
values 
decrease, the cosmological time is assumed to increase.}}}
\label{plot2}
\end{figure}

Having the scalar potential (\ref{scalarpotfnt1}) at hand, we can qualitatively 
investigate when the slow-roll approximation is violated. Practically this occurs when 
the 
slow-roll parameter $\epsilon$ becomes $\epsilon \gtrsim 1$. In the case of a 
scalar-tensor theory with two canonical scalar fields, the slow-roll parameter $\epsilon$ 
is equal to 
\cite{Starobinsky:1982ee,Starobinsky:1982ee1,KofmanStarobinsky,Lyth:2005fi,
Vernizzi:2006ve,
Choi:2007su,Alabidi:2006qa,Alabidi:2005qi,Sasaki:1998ug}:
\begin{equation}
\label{slowrollepsilon}
\epsilon (\varphi,\sigma)=\epsilon_{\varphi}+\epsilon_{\sigma}\, ,
\end{equation} 
with $\epsilon_{\varphi}$ and $\epsilon_{\sigma}$ reading as
\begin{equation}
\label{newparameters}
\epsilon_{\varphi}=\frac{1}{2\kappa^2}\left[\frac{\partial_{\varphi}V(\varphi,\sigma)}{
V(\varphi, 
\sigma)}\right]^2,{\,}{\,}{\,}\epsilon_{\sigma}=\frac{1}{2\kappa^2}\left[\frac{\partial_{
\sigma}V(\varphi,\sigma)}{V(\varphi, \sigma)}\right]^2\, .
\end{equation}
Thus, by using the potential (\ref{scalarpotfnt1}) we may directly evaluate the 
slow-roll parameter $\epsilon $ and study it's evolution as a function of the two scalar 
fields. Assuming again large field values for small cosmological times, which decrease as 
the cosmic time approaches the end of inflation, in Fig. \ref{plot2} we present the 
behavior of the slow-roll parameter $\epsilon$ as a function of the scalar fields.
As we observe the slow-roll approximation is violated for a combination of values of the 
two fields, as expected. 

Before closing this subsection, we shall investigate the impact of the singularities on 
the cosmological parameters of the two-field model. In particular, we are interested in 
the cases where the singularities may lead to inconsistencies. As we mentioned earlier, 
the main inconsistency is traced in the definition of the $e$-fold number $N$, which is 
defined as
\begin{equation}
\label{nefold1}
N=-\int_{t_f}^{t}H(t_*)\mathrm{d}t\, ,
\end{equation}
where $t_*$ is the Hubble exit during inflation. The final time $t_f$ corresponds to a 
uniform-density hypersurface, at which inflation is considered to end. By taking into 
account (\ref{hubtwonew}) and (\ref{deffunctions}), the $e-$folding number writes as
\begin{equation}
\label{typeivconsistent}
N=\frac{(\nu_1+\nu_2) \left[(t_1-t_f)^{\alpha+1 
}-(-t_*+t_1)^{\alpha+1 }\right]}{\alpha+1}\, .
\end{equation}
Hence, it is obvious that when $\alpha<-1$ the above expression is always singular when 
$t_f=t_1$, due to the last term. However, if the singularity is of Type IV there is no 
inconsistency, which is the main point of this subsection.

Finally, as we mentioned earlier too, we note that the complete analysis of the 
$e$-folding number for the two-field case can be quite complicated, and a detailed 
investigation would require the use of the $\delta$N approximation. However, such a study 
stretches beyond the scope of the present work, and the reader is referred to 
\cite{Starobinsky:1982ee,Starobinsky:1982ee1,KofmanStarobinsky,Lyth:2005fi,
Vernizzi:2006ve,Choi:2007su,
Alabidi:2006qa,Alabidi:2005qi,Sasaki:1998ug}
and references therein.


\subsection{Singular inflation with singular dark energy}
\label{subsingDE}

In this subsection we are interested in constructing a cosmological scenario where 
singular inflation gives rise to radiation and matter eras, and eventually to a dark 
energy epoch that results in a Big-Rip singularity. In order to succeed this, and 
following the previous subsections, we use two scalar fields, and thus this is another 
indication of the capabilities of such a model. 

Having in mind the discussion on the singularity types of the Hubble rate (\ref{IV1}), we 
deduce that a Hubble rate that captures the above features, namely a singular inflation 
with a type IV singularity and a dark energy epoch resulting to a Big Rip, writes as
\be
\label{IV1anojsingDE}
H(t) =c_0+b_0 \left( t_s - t \right)^\alpha +d_0 \left( t_B - t \right)^\gamma\, ,
\ee
where $\alpha>1$ and $\gamma<-1$, with $c_0$, $b_0$, $d_0$ constants, and where $t_s$ 
and $t_B$ are respectively the time of the inflationary type IV singularity and of the 
Big Rip. We need to note that $\alpha$ is assumed to take values of (\ref{IV2noj}), 
therefore complex values for the Hubble rate are avoided for the physically interesting 
interval $t_s<t<t_B$. 
Practically, since the denominator of the 
fraction 
appearing in  (\ref{IV2noj}) is an odd integer always, then this procedure is always 
valid. Definitely, there are extra complex conjugate 
branches too, but we only focus on the real negative one.

Assuming as usual that the two scalar fields $\phi$ and $\chi$ depend only on 
the cosmic-time 
coordinate $t$, and also that the spacetime is described by a flat FRW metric of the form 
(\ref{JGRG14}), we extract the Friedmann equations in the form
\begin{eqnarray}
\label{A2singDE}
&&\omega(\phi) {\dot \phi}^2 + \eta(\chi) {\dot \chi}^2
= - \frac{2}{\kappa^2}\dot H\, ,\\
&&
V(\phi,\chi)=\frac{1}{\kappa^2}\left(3H^2 + \dot H\right)\, .
\label{A2bbsingDE}
\end{eqnarray}
Following the procedure of the previous subsections we impose the parametrization 
\begin{eqnarray}
\label{A3singDE}
&&\omega(t) + \eta(t)=- \frac{2}{\kappa^2}f'(t)\, , \nonumber\\
&&
V(t)=\frac{1}{\kappa^2}\left[3f(t)^2 + f'(t)\right]\, ,
\end{eqnarray}
which is consistent with the explicit solution of (\ref{A2singDE}) and (\ref{A2bbsingDE}):
\be
\label{A4singDE}
\phi=\chi=t\, ,\quad H=f(t)\, .
\ee
A convenient choice for the kinetic functions would be  
\begin{eqnarray}
&&\omega(\phi) =-\frac{2}{\kappa^2}\left\{f'(\phi)
- \sqrt{\alpha_1(\phi)^2 + f'(\phi)^2} \right\}>0\, , \nonumber\\
&&
\eta(\chi) = -\frac{2}{\kappa^2}\sqrt{\alpha_1(\chi)^2 + f'(\chi)^2}<0\, ,
\label{A5singDE}
\end{eqnarray}
with  $\alpha_1(x)$ an arbitrary function of its argument. It proves useful to define 
a new function $\tilde f(\phi,\chi)$ through
\be
\label{A6singDE}
\tilde f(\phi,\chi)\equiv - \frac{\kappa^2}{2}\left[\int d\phi 
\omega(\phi) + \int d\chi \eta(\chi)\right]\,,
\ee
which has the characteristic property $\tilde f(t)=f(t)$ which fixes the constants of 
integration in (\ref{A6singDE}). Assuming that 
$V(\phi,\chi)$ is given in terms of the function $\tilde f(\phi,\chi)$ as
\be
\label{A8singDE}
V(\phi,\chi)=\frac{1}{\kappa^2}\left[3{\tilde f(\phi,\chi)}^2
+ \frac{\partial \tilde f(\phi,\chi)}{\partial \phi}
+ \frac{\partial \tilde f(\phi,\chi)}{\partial \chi} \right]\, ,
\ee
we find that in addition to the Friedmann equations given in 
(\ref{A2singDE}) and (\ref{A2bbsingDE}) we have
\begin{eqnarray}
\label{A9singDE}
&&0 = \omega(\phi)\ddot\phi + \frac{1}{2}\omega'(\phi) {\dot \phi}^2
+ 3H\omega(\phi)\dot\phi + \frac{\partial \tilde V(\phi,\chi)}{\partial 
\phi} \, , \\
&&
0 = \eta(\chi)\ddot\chi + \frac{1}{2}\eta'(\chi) {\dot \chi}^2
+ 3H\eta(\chi)\dot\chi + \frac{\partial \tilde V(\phi,\chi)}{\partial 
\chi}\, .
\label{A9bbsingDE}
\end{eqnarray}
Hence, using the above kinetic functions $\omega (\phi)$, $\eta (\chi)$ 
and the scalar potential $V(\phi,\chi)$ we obtain  a two-field scalar-tensor 
model with cosmological evolution of the form (\ref{IV1anojsingDE}).
 
As a next step we need to impose an ansatz for the function $\alpha_1(x)$ 
appeared in (\ref{A5singDE}). We choose it as
\begin{equation}
\label{alphadefsingDE}
\alpha_1 (x)=\sqrt{b_0\alpha(-x +t_s)^{\alpha -1}-H'(x)^2}\, ,
\end{equation}
and therefore (\ref{A5singDE}) give  
\begin{eqnarray}
&&\omega (\phi )=\frac{2 \left[b_0 \alpha  (t_s-\phi )^{\alpha-1}+d_0 \gamma  
(t_B-\phi )^{\gamma-1}+\sqrt{b_0 \alpha  
(t_s-\phi )^{\alpha-1}}\right]}{\kappa ^2}\, ,\nonumber\\
&&
\eta (\chi)= -\frac{2 \sqrt{b_0 \alpha  (t_s-\chi )^{-1+\alpha }}}{\kappa ^2}\, .
\label{omeganforhilltopdoublesingDE}
\end{eqnarray}
As expected, one of the fields is canonical while the other one is ghost. The 
corresponding function $\tilde f(\phi,\chi)$ from (\ref{A6singDE}) becomes
\begin{equation}
\label{barfsingDE}
\tilde f(\phi,\chi)=- b_0  (t_s-\phi )^{\alpha }-d_0  (t_B-\phi )^\gamma  
-\frac{2 \sqrt{
b_0 \alpha  } (t_s-\phi )^{\frac{\alpha+1 }{2}} }{(\alpha +1) }
-\frac{2 \sqrt{
b_0 \alpha  } (t_s-\chi )^{\frac{\alpha+1 }{2}} }{(\alpha +1) }\, ,
\end{equation}
and thus the scalar potential $V(\phi,\chi)$ from (\ref{A8singDE}) becomes
\begin{align}
\label{scalarpotentialhilltopsingDE}
V(\phi,\chi)=& \frac{b_0 \alpha    (t_s-\phi )^{\alpha-1}+d_0 \gamma(t_B-\phi 
)^{\gamma-1}}{\kappa^2} \nonumber\\
& +\frac{2 \sqrt{b_0 \alpha 
}(t_s-\phi 
)^{\frac{\alpha-1 }{2}}+\sqrt{b_0 \alpha } (\alpha-1 )\text{  }(t_s-\phi 
)^{\frac{\alpha-1}{2} 
} }{(1+\alpha )\kappa ^2}
\nn
&
+  \left[\frac{2 \sqrt{b_0 \alpha  } (t_s-\chi )^{\frac{\alpha-1}{2}}}{(1+\alpha ) \kappa 
^2}-\frac{  \sqrt{b_0 \alpha }(\alpha-1 ) (t_s-\chi 
)^{\alpha-1}}{(1+\alpha ) \kappa ^2}\right]
\nn
&
+\frac{3}{\kappa^2}   \left\{ b_0  (t_s-\phi )^{\alpha }+d_0  (t_B-\phi )^\gamma  
+\frac{2 \sqrt{
b_0 \alpha  } (t_s-\phi )^{\frac{\alpha+1 }{2}} }{(\alpha +1)  }
+\frac{2 \sqrt{
b_0 \alpha  } (t_s-\chi )^{\frac{\alpha+1 }{2}} }{(\alpha+1 )  }\right\}^2\, .
\end{align}
  
In summary, the above two-field model can give rise to the Hubble rate 
(\ref{IV1anojsingDE}), which corresponds to a universe with a singular (of type IV) 
inflation, resulting in a dark-energy phase that ends with a Big Rip.

\subsection{Barrow's two-field model}
\label{subBarrow}

In this subsection, for completeness, we present the realization of the singular 
inflation in the Barrow's two-field model \cite{Barrow:2015ora}. This model can exhibit 
the Type IV singularity consistently incorporated in it's scalar-tensor theoretical 
framework. In particular, we consider a Hubble rate of the form
\be
\label{IV1bar}
H(t) = b_0 \left( t_s - t \right)^\alpha\, .
\ee
We choose the arbitrary function $\alpha_1(x)$ appearing in (\ref{A5}) to be 
\begin{equation}
\label{alphadefbar}
\alpha_1(x)=\sqrt{2 \alpha  b_0}(-x +t_s)^{(\alpha -1)/2}\, ,
\end{equation}
and substituting this into (\ref{A5}) we can easily obtain the corresponding kinetic 
functions $\omega (\phi)$ and $\eta (\chi)$ as
\begin{eqnarray}
&&\omega (\phi )=-\frac{2 \left[-b_0 \alpha  (t_s-\phi )^{-1+\alpha }-\sqrt{b_0 \alpha  } 
(t_s-\phi )
^{\frac{-1+\alpha }{2}}\right]}{\kappa ^2} \, ,
\nonumber\\
&&
\eta (\chi)= -\frac{2 \sqrt{b_0 \alpha  } (t_s-\phi )^{\frac{-1+\alpha }{2}}}{\kappa ^2} 
\, .
\label{omeganforhilltopdoublebar}
\end{eqnarray}
In the case at hand, and assuming that $\alpha$ has the form (\ref{IV2}), namely 
$\alpha= n/(2m + 1)$, with $n$ constrained to be an even integer and $\alpha>1$, the 
kinetic function $\omega(\phi)$ is rendered positive, while the kinetic function 
$\eta(\chi)$ becomes always negative. Thus, we can promptly find the function $\tilde 
f(\phi,\chi)$, which in this case is 
\begin{equation}
\label{barfbar}
\tilde f(\phi,\chi)=-\frac{1}{2} \kappa ^2 \left\{-\frac{2 \left[-b_0 (1+\alpha ) 
(t_s-\phi 
)^{\alpha }-2 \sqrt{b_0 \alpha  } (t_s-\phi )^{\frac{1+\alpha }{2}}\right]}{(1+\alpha ) 
\kappa 
^2}+\frac{4 \sqrt{b_0 \alpha  }(t_s-\chi )^{\frac{1+\alpha }{2}}}{(1+\alpha ) \kappa 
^2}\right\}\, ,
\end{equation}
and consequently the corresponding scalar potential $V(\phi,\chi)$ reads
\begin{align}
\label{scalarpotentialhilltopbar}
V(\phi,\chi)=&\frac{b_0 \alpha  (1+\alpha ) (t_s-\phi )^{-1+\alpha }+\sqrt{b_0 \alpha } 
(1+\alpha )
 (t_s-\phi )^{\frac{1}{2} (-1+\alpha )}}{(1+\alpha ) \kappa ^2} +\frac{\sqrt{b_0 
\alpha 
}}{\kappa ^2} (t_s-\chi )^{\frac{1}{2} (-1+\alpha )}
 \nn 
 & +\frac{3}{4} \kappa ^2 \left\{-\frac{2 \left[-b_0 (1+\alpha ) (t_s-\phi )^{\alpha }-2 
\sqrt{b_0 \alpha } (t_s-\phi )^{\frac{1+\alpha }{2}}\right]}{(1+\alpha ) \kappa 
^2}+\frac{4 \sqrt{b_0 \alpha } 
(t_s-\chi )^{\frac{1+\alpha }{2}}}{(1+\alpha ) \kappa ^2}\right\}^2 \, .
\end{align}
Therefore, the two-field scalar-tensor theory that generates the cosmological evolution 
of (\ref{IV1bar}), is given by expressions (\ref{omeganforhilltopdoublebar}) and 
(\ref{scalarpotentialhilltopbar}).

\subsection{The case of multiple scalar fields}
\label{subsmultiple}

We close the current section by presenting,  for completeness, the possible 
generalization of the above results in the case of multiple fields. Such a framework 
offers huge capabilities and possibilities of cosmological evolutions. We provide here 
the general framework of this extension following \cite{sergnoj} (see also 
\cite{Setare:2008si}). The scalar-tensor action in the case of $n$ real scalar fields 
$\phi_i$, $i=1,..n$, reads
\be
\label{A1gener}
S=\int d^4 x \sqrt{-g}\left\{\frac{1}{2\kappa^2}R
 - \sum_{i=1}^{n}\frac{1}{2}\omega(\phi_i)\partial_\mu \phi_i \partial^\mu \phi_i
 - V(\phi_i)\right\}\, .
\ee
Assuming a flat FRW background, and assuming that the scalar fields are functions only of 
the cosmic time $t$, the corresponding Friedmann equations write as
\begin{eqnarray}
\label{A2gener}
&&\sum_{i=1}^n\omega(\phi_i) {\dot \phi}^2
= - \frac{2}{\kappa^2}\dot H \, , \\
&&
V(\phi_i)=\frac{1}{\kappa^2}\left(3H^2 \right)-\frac{1}{2}\sum_{i=1}^n\omega 
(\phi_i)\dot{\phi_i}^2\, .
\end{eqnarray}
We choose the function $f(t)$ in order to satisfy the following equation: 
\be
\label{A3gener}
\sum_{i=1}^n\omega_i(t)=- \frac{2}{\kappa^2}f'(t)\, ,
\ee
and therefore we obtain
\begin{align}
\label{neweqns}
& 
\omega_i(\phi_i)=-\frac{2}{\kappa^2}\frac{\mathrm{d}f(\phi_1,\phi_2,...,\phi_n)}{\mathrm{d
}\phi_i} \, ,
\nn &
V(\phi_1,\phi_2,...,\phi_n)=\frac{1}{\kappa^2}\left [ 
3f(\phi_1,\phi_2,...,\phi_n)^2+\sum_{i=1}^n\frac{\mathrm{d}f(\phi_1,\phi_2,...,\phi_n)}{
\mathrm{d}\phi_i}\right ] \, ,
\end{align}
with the function $f(t_i)=f(t,t,...,t)=f(t)$. The solution of the reconstruction method 
is then given by
\be
\label{A4gener}
\phi=\chi=t\, ,\quad H=f(t)\, .
\ee
By appropriately choosing some arbitrary functions $g_n(\phi_i)$ we can make the kinetic 
functions $\omega_i(\phi_i)$ to be  
\begin{eqnarray}
\label{A5gener}
&&\omega_1(\phi_1) =-\frac{2}{\kappa^2}\left\{f'(\phi_1)+g_2(\phi_1)+\cdots+g_n(\phi_1) 
\right\}\, , \nonumber
\\
&&\omega_2(\phi_2)= \frac{2}{\kappa^2}g_2(\phi_2)\, ,\nonumber\\
&&\ \ \ \ \ \ \ \ \  
\vdots
\nonumber\\
&&\omega_n(\phi_n)=\frac{2}{\kappa^2}g_n(\phi_n)\, .
\end{eqnarray}
Hence, one may choose to have ghost and non-ghost scalar fields and describe the 
cosmological evolution in a desired way. We do not analyze this case in more details 
since it lies beyond the scope of the present work.

\section{$F(R)$-gravity analysis}
\label{FRsection}

As a last section of this work, and for completeness, we will investigate the case of 
$F(R)$ gravity 
\cite{Starobinsky:1980te,Nojiri:2006ri,Capozziello:2010zz,Capozziello:2011et,
delaCruzDombriz:2012xy,Carroll:2003wy,Capozziello:2002rd,Capozziello:2005ku,reviews1}, 
which is closely related to scalar-field theory, focusing again on the Type IV 
realization 
of a Hubble rate of the form (\ref{IV}). We will follow the reconstruction procedure of 
\cite{Nojiri:2006gh,Capozziello:2006dj,Nojiri:2006be} (see also \cite{sergbam08}). 

We start by recalling that in the case where $\alpha\gg 1$, the scalar-tensor potential 
which perfectly describes this cosmology is a hilltop-like potential of the form 
(\ref{hilltop}). We shall assume that the parameters $c_0$ and $b_0$ 
appearing in (\ref{IV1}) are chosen in order to satisfy $c_0\ll 1$ and 
$b_0\ll 1$, and in addition $\alpha\gg 1$. Furthermore, we remind that for an FRW metric, 
the Ricci scalar writes as
\begin{equation}
\label{ricciscal}
R=6(2H^2+\dot{H})\, .
\end{equation}
As the cosmic time $t$ approaches the finite time $t_s$, then $t-t_s\rightarrow 0$. We 
shall use this property later on in this section, in order to find approximate 
expressions 
for the $F(R)$ gravity. In addition, as $t\rightarrow t_s$, the Ricci scalar approaches 
the limiting value $R=12c_0^2$, as can be seen by calculating it's analytic form 
\begin{equation}
\label{desitterricci}
R=12 c_0^2+24 b_0 c_0 (-t+t_s)^{\alpha }+12 b_0^2 (-t+t_s)^{2 \alpha }-6 b_0 
(-t+t_s)^{-1+\alpha } \alpha \, .
\end{equation} 
We shall further investigate the physical consequences of our assumptions in the end of 
this 
section.

The Jordan frame action of a pure $F(R)$ gravity is  
\begin{equation}
\label{action1dse}
\mathcal{S}=\frac{1}{2\kappa^2}\int \mathrm{d}^4x\sqrt{-g}F(R)\, ,
\end{equation}
with pure referring to the absence of matter fluids. Varying action (\ref{action1dse}) 
with respect to the metric tensor leads to the Friedmann equation
\begin{equation}
\label{frwf1}
 -18\left [ 4H(t)^2\dot{H}(t)+H(t)\ddot{H}(t)\right ]F''(R)+3\left [H^2(t)+\dot{H}(t) 
\right ]F'(R)-
\frac{F(R)}{2}=0\, .
\end{equation}
By making use of an auxiliary scalar field $\phi $ (which has no kinetic term), the 
$F(R)$ gravity action of (\ref{action1dse}) can be rewritten as 
\begin{equation}
\label{neweqn123}
S=\int \mathrm{d}^4x\sqrt{-g}\left [ P(\phi )R+Q(\phi ) \right ]\, .
\end{equation}
By finding the exact analytic form of the functions $P(\phi )$ and $Q(\phi )$, which have 
an explicit dependence on the scalar field $\phi$, we may easily obtain the specific 
$F(R)$ gravity. This can be done by varying the action of (\ref{neweqn123}) with 
respect to the auxiliary scalar degree of freedom $\phi$. Such a variation gives
\begin{equation}
\label{auxiliaryeqns}
P'(\phi )R+Q'(\phi )=0\, ,
\end{equation}
with the prime now denoting differentiation with respect to $\phi$. Solving 
Eq.~(\ref{auxiliaryeqns}) with respect to $\phi $ will provide us with the function 
$\phi (R)$, and then the $F(R)$ gravity can be easily found by substituting $\phi (R)$ to 
the auxiliary $F(R)$ action of (\ref{neweqn123}). Hence, the reconstructed $F(R)$ gravity 
will be
\begin{equation}
\label{r1}
F(\phi( R))= P (\phi (R))R+Q (\phi (R))\, .
\end{equation}

Let us apply the above reconstruction procedure for the case where the Hubble rate is
\begin{equation}
\label{hurnew}
H(t)=c_0+b_0\left( -t+t_s \right)^{\alpha}\, .
\end{equation}
By varying action (\ref{neweqn123}) with respect to the metric, we obtain the 
equations
\begin{align}
\label{r2}
0= & -6H^2P(\phi (t))-Q(\phi (t) )-6H\frac{\mathrm{d}P\left (\phi (t)\right 
)}{\mathrm{d}t}=0\, , \nn
0=& \left ( 4\dot{H}+6H^2 \right ) P(\phi (t))+Q(\phi (t) )+2\frac{\mathrm{d}^2P(\phi 
(t))}{\mathrm 
{d}t^2}+\frac{\mathrm{d}P(\phi (t))}{\mathrm{d}t}=0\, .
\end{align}
Eliminating $Q(\phi (t))$ from (\ref{r2}) results to
\begin{equation}
\label{r3}
2\frac{\mathrm{d}^2P(\phi (t))}{\mathrm {d}t^2}-2H(t)\frac{\mathrm{d}P(\phi 
(t))}{\mathrm{d}t}+4\dot{H}P(\phi (t))=0\, .
\end{equation}
In this way, given the Hubble rate, the explicit form of $P(t)$ can be found and from 
this we can obtain $Q(t)$ accordingly. We mention that, due to fact that the $F(R)$ 
action (\ref{action1dse}) is mathematically equivalent to the auxiliary $F(R)$ action 
(\ref{neweqn123}), the auxiliary scalar field is identical to the cosmic time $t$, that 
is 
$\phi =t$. This was proven in detail in the Appendix A of Ref.~\cite{Nojiri:2006gh}.
Substituting the Hubble rate (\ref{hurnew}) in (\ref{r3}), we obtain the following 
second-order differential equation:
\begin{equation}
\label{ptdiffeqn}
2\frac{\mathrm{d}^2P(t)}{\mathrm {d}t^2}-2\left[ c_0+b_0(-t+t_s)^{\alpha 
}\right]\frac{\mathrm{d}P(
t)}{\mathrm{d}t}-4 b_0 (-t+t_s)^{-1+\alpha } \alpha P(t)=0\, ,
\end{equation}
which by setting $x=-t+t_s$ can be rewritten as  
\begin{equation}
\label{dgfere}
2\frac{\mathrm{d}^2P(x)}{\mathrm {d}x^2}-2\left( c_0+b_0x^{\alpha 
}\right)\frac{\mathrm{d}P(x)}{\mathrm{d}x}-2 b_0 x^{-1+\alpha } \alpha P(x)=0\, .
\end{equation}
The general solution of the above equation is difficult to be extracted, thus we will  
follow approximations. In particular, since $x$ is small the differential equation 
becomes
\begin{equation}
\label{dgferemod}
2\frac{\mathrm{d}^2P(x)}{\mathrm {d}x^2}-2c_0\frac{\mathrm{d}P(x)}{\mathrm{d}x}=0\, ,
\end{equation}
with solution
\begin{equation}
\label{genrealsol}
P(x)= \frac{ \e^{c_0 x}  c_1}{1+c_0}+c_2\, ,
\end{equation}
with $c_1$, $c_2$ being arbitrary constants. Then, the function $Q(x)$ can easily be 
calculated using (\ref{r2}) as
\begin{align}
\label{qtanalyticform}
Q(x) =-6 c_0^2 c_2-\frac{12 c_0^2 c_1 \e^{c_0 x}}{1+c_0}-12 b_0 c_0 c_2 x^{\alpha 
}-\frac{18 b_0 c_
0 c_1 \e^{c_0 x} x^{\alpha }}{1+c_0}-6 b_0^2 c_2 x^{2 \alpha }-\frac{6 b_0^2 c_1 \e^{c_0 
x} x^{2 \alpha }}{1+c_0}\, .
\end{align}
Substituting the final forms of the functions $P(x)$ and $Q(x)$ into 
(\ref{auxiliaryeqns}), and by solving with respect to $x$, we can find the function 
$x(R)$. For the case $t\rightarrow t_s$ that we are interested in, i.e. for 
$x\rightarrow 0$, we can expand the involved $P'(x)$ and $Q'(x)$. 
Doing so we acquire
\begin{equation}
\label{pseries}
P'(x)\simeq  \frac{c_0 c_1}{1+c_0}+\frac{c_0^2 c_1 x}{1+c_0}+\frac{c_0^3 c_1 x^2}{2 
(1+c_0)}\, ,
\end{equation}
and  
\begin{align}
\label{qsapprox1}
Q'(x)\simeq&  x^{2 \alpha } \left[-\frac{6 \left(b_0^2 c_0 c_1\right)}{1+c_0}-\frac{6 
\left(b_0^2 
c_0^2 c_1\right) x}{1+c_0}\right]+x^{\alpha } \left[-\frac{12 \left(b_0 c_0^2 
c_1\right)}{1+c_0}-\frac{12 \left(b_0 c_0^3 c_1\right) x}{1+c_0}\right]
\nn 
&+x^{\alpha } 
\left[-\frac{6 \left(b_0 c_0^2 
c_1\right)}{1+c_0}-\frac{6 \left(b_0 c_0^3 c_1\right) x}{1+c_0}\right]+\left[-\frac{12 
\left(c_0^3 
c_1\right)}{1+c_0}-\frac{12 \left(c_0^4 c_1\right) x}{1+c_0}\right]
\nn 
&+x^{\alpha } 
\left[-\frac{6 (b_0 
c_0 c_1 \alpha )}{(1+c_0) x}-\frac{6 \left(b_0 c_0^2 c_1 \alpha \right)}{1+c_0}\right]
\nn 
& +x^{2 \alpha } \left\{-\frac{12 \left[b_0^2 
\left(\frac{c_1}{1+c_0}+c_2\right) \alpha 
\right]}{x}-\frac{12 \left(b_0^2 c_0 c_1 \alpha \right)}{1+c_0}\right\}
\nn 
&+x^{\alpha } 
\left\{-\frac{12 \left[b_0 c_0 \left(\frac{c_1}{1+c_0}+c_2\right] \alpha 
\right)}{x}-\frac{12 \left(b_0 c_0^2 c_1 \alpha \right)}{1+c_0}\right\}\, .
\end{align}
Since $\alpha\gg 1$, keeping the most dominant terms in (\ref{qsapprox1}) we 
obtain
\begin{align}
\label{qsapprox}
& Q'(x)\simeq \left(-\frac{12 c_0^3 c_1}{1+c_0}-\frac{12 c_0^4 
c_1 x}{1+c_
0}\right) \, .
\end{align}
Hence, substituting (\ref{pseries}) and (\ref{qsapprox}) into (\ref{auxiliaryeqns}) we 
acquire
\begin{equation}
\label{finalxr}
x\simeq \frac{12 c_0^3-c_0 R+\sqrt{144 c_0^6-c_0^2 R^2}}{c_0^2 R} \, .
\end{equation}
Thus, the $F(R)$ gravity form that can reproduce the Type IV singularity, can be obtained 
straightforwardly upon substitution of (\ref{finalxr}) into (\ref{r1}), and reads as
\begin{equation}
\label{finalfrgravity}
F(R)\simeq \frac{144 c_0^4 c_1}{(1+c_0) R}-\frac{144 c_0^5 c_1}{(1+c_0) R}+c_2 R+\frac{12 
c_0 c_1 \sqrt{144 c_0^6-c_0^2 R^2}}{(1+c_0) R}-\frac{12 c_0^2 c_1 \sqrt{144 c_0^6-c_0^2 
R^2}}{(1+c_0) R}\, .
\end{equation}
Observing the above form, we deduce that by appropriately choosing the arbitrary constant 
$c_2$ we can make the $F(R)$ function to be the usual Einstein-Hilbert gravity plus 
curvature corrections. Indeed, for $c_2=1$, the $F(R)$ form becomes
\begin{equation}
\label{finalfrgravityprofinal}
F(R)\simeq  R+\frac{144 c_0^4 c_1}{(1+c_0) R}-\frac{144 c_0^5 c_1}{(1+c_0) R}+\frac{12 
c_0 
c_1 \sqrt{144 c_0^6-c_0^2 R^2}}{(1+c_0) R}-\frac{12 c_0^2 c_1 \sqrt{144 c_0^6-c_0^2 
R^2}}{(1+c_0) R}\, .
\end{equation}
We can further simplify the final form of the $F(R)$ gravity by recalling that 
(\ref{finalfrgravityprofinal}) is valid only near the Type IV singularity, in which case 
the Ricci scalar approaches the limiting value $R=R_{\mathrm{c}}=12 c_0^2$ (see 
(\ref{desitterricci})). Therefore, for $R\rightarrow R_c$, the last two terms 
of (\ref{finalfrgravityprofinal}) are nearly zero. Consequently, the final form of the 
$F(R)$ gravity is approximately  
\begin{equation}
\label{finalfrgravityafterm}
F(R)\simeq R+\frac{144 c_0^4 c_1\left(1-c_0\right)}{(1+c_0)}\frac{1}{R}\, .
\end{equation}
This type of $F(R)$ gravity was described in \cite{Carroll:2003wy}. 
Recall our initial assumptions, namely $c_0\ll 1$ and $b_0\ll 1$, and in addition 
$\alpha\gg 1$. Furthermore, as $t\rightarrow t_s$ the scalar curvature approaches the 
limiting 
value $R\rightarrow R_c$. Therefore, we have an appealing physical 
picture if we additionally require that $t_s$ is a future late-time singularity. Indeed, 
the $F(R)$ gravity is of the form $F(R)\sim R+\frac{C}{R}$ and the scalar curvature 
approaches a late-time point, which since $c_0\ll 1$ this limiting curvature value can 
be chosen to have a very small value. Hence, the $F(R)$ form of 
(\ref{finalfrgravityafterm}) can successfully describe late-time acceleration near the 
future Type IV singularity, which was the goal of the present section. 

We mention that the $F(R)$ gravity of the form (\ref{finalfrgravityafterm}) can
consistently describe dark energy and late-time acceleration, as was demonstrated in 
\cite{Carroll:2003wy}, and with the presence of an $R^2$ term it can unify inflation with 
dark energy \cite{nojiri2003a}. However, the model 
(\ref{finalfrgravityafterm}) is not completely in agreement with observations, having 
problems with the correct description of the gravitational interaction in the Solar 
System. Nevertheless, the model (\ref{finalfrgravityafterm}) turns out to be a very 
useful 
approximate form of consistent and realistic versions of $F(R)$ gravity (see 
\cite{reviews1} for a review). However, we have to note that the de Sitter solution for 
this model is $R_{\mathrm{ds}}=12c_0^2\sqrt{-3c_1}$, thus for negative $c_1$ the 
model we found may not describe dark energy, as was shown in \cite{ame1,ame2}, since 
$F''(R)<0$. Additionally, as it was stressed in \cite{staro}, in order 
for the scalaron mass to be less than the Planck mass, the  $F''(R)$ has to be bounded 
for a finite $R$, which is not satisfied by (\ref{finalfrgravity}).

\section{Conclusions}
\label{conclus}

In this work we demonstrated that finite time singularities of Type IV, can consistently 
be incorporated in the Universe's cosmological evolution, either appearing in the 
inflationary era, or in the late-time regime. We thoroughly discussed the most 
consistent framework in which these can be realized without instabilities. In 
particular, using only one scalar field then the linear-level instabilities can in 
principle occur at the cosmological time at which the phantom divide is crossed during 
the dynamical evolution, as we showed to happen in the case of a hilltop-like potential. 
On the other hand, when two fields are involved, one being canonical and one being 
ghost, it is possible to avoid such instabilities during the phantom-divide crossing. 
In 
addition, despite their increased level of complexity, the two-field scalar-tensor 
theories prove to be able to offer a plethora of possible viable cosmological scenarios, 
at which various types of cosmological singularities can be realized. For instance, it is 
possible to describe phantom inflation with the appearance of a Type IV singularity, and 
phantom late-time acceleration which ends in a Big Rip singularity.

For completeness, we also presented the Type IV realization in the context of $F(R)$ 
gravity, which is known to be connected with a scalar-field theory through conformal 
transformations. In particular, we showed that specifically reconstructed $F(R)$ 
forms can generate cosmological scenarios with a dynamical evolution corresponding to 
hilltop-like potentials. As we evinced, the resulting $F(R)$ gravity is of the form 
$R+\frac{A}{R}$, and this can be quite physically appealing if the Type IV singularity 
occurs at late times, since the resulting model is known to produce late-time 
acceleration. 

In the absence of a complete understanding of the nature of cosmological singularities, 
we believe that with the present study we demonstrated that not all singularities can be 
harmful, at least for the cases where observational consistency is the main goal of a 
self-consistent physical theory. Indeed, even in observable quantities that contain 
higher derivatives of the scale factor, the effect of singularities, and particularly of 
Type IV singularities, might be negligible, as for example in the jerk parameter. 
Definitely, a lot of effort is needed to theoretically understand the singularities, but 
probably these Type IV singularities might be a classical bridge between classical 
cosmology and the ultimate quantum gravitational cosmology, which offers a complete 
description or at least a remedy of these classically unwanted spacetime points. Some 
steps towards these directions of research are offered by loop quantum gravity, in which 
no singularities occur 
\cite{LQC,Bojowald:2008ik,Quintin:2014oea,deHaro:2012xj,Cai:2011zx}. 
However, we mention that it is possible even at the classical level to remedy 
singularities \cite{Nojiri:2005sx} or even avoid them \cite{Nojiri:2004ip}, when quantum 
corrections are taken into account, as in \cite{Elizalde:2004mq,Nojiri:2005sx}, and thus 
we hope to address this issue in the future in more details. 

It would be interesting to notice that phantom-non-phantom transitions as well as 
non-singular evolution appear also in the case of bouncing cosmology. Hence, one could 
follow the procedures and the techniques of the present work, in order to investigate the 
effects of Type IV singularities, and probably of Type III ones, in the context of 
various ekpyrotic \cite{Khoury:2001wf,ekp,Lehners:2008vx} or bounce Universe scenarios 
\cite{Starobinskii1978,Novello:2008ra,Brandenberger:1993ef,bounce3,bounce5,
Qiu:2013eoa,Cai:2014xxa,Khoury:2001zk,Erickson:2003zm, 
Lehners:2013cka,Saridakis:2007cf,Creminelli:2007aq,Cai:2010zma, HLbounce,
Bojowald:2001xe,Martin:2003sf,Cai:2012ag,Ashtekar:2007tv,
Cailleteau:2012fy,Haro:2014wha,Amoros:2014tha,WilsonEwing:2012pu}. This thorough analysis 
lies beyond the scope of the present work and will be addressed elsewhere.

Finally, we have to mention that in this work we did not address in general the issue of 
possible analyticity loss beyond $\varphi=0$. As it is also commented by Barrow and 
Graham in \cite{Barrow:2015ora}, for large-field inflation potentials of the form 
$\sim\varphi^n$, there are fractional powers of $n$ for which the potential is always 
real, and extensibility through $\varphi=0$ can be achieved leading to a Big Crunch (for 
instance when the potential becomes negative). The forms of the potentials we used are 
exactly of this class (large-field inflation ones), or hilltop-like potentials. 
Nevertheless, in general, the analyticity issue around $\varphi=0$ is very interesting, 
both from mathematical as well as physical point of view. This issue however deserves a 
detailed investigation of its own and is left for a future project.

\begin{acknowledgments}
This work is supported  in part by the JSPS Grant-in-Aid for Scientific Research (S) \# 
22224003 and (C) \# 23540296 (S.N.) and in part by MINECO (Spain), projects FIS2010-15640 
and FIS2013-44881. The research of ENS is implemented within the framework of the Action 
``Supporting Postdoctoral Researchers'' of the Operational Program ``Education and 
Lifelong Learning'' (Actions Beneficiary: General Secretariat for Research and 
Technology), and is co-financed by the European Social Fund (ESF) and the Greek State. 
\end{acknowledgments}


\end{document}